\documentclass[preprint,12pt]{elsarticle}
\usepackage{amsmath}
\usepackage{amsfonts}
\usepackage{amssymb}
\usepackage[utf8]{inputenc}
\usepackage[T2A]{fontenc}
\usepackage{bm}
\usepackage{graphicx}

\begin{document}

\title{General solution \textit{vs}  spin invariant eigenstates \\
of the Dirac equation with the Coulomb potential  \\
}

\author{L.S. Brizhik$^1$, A.A. Eremko$^1$, V.M. Loktev$^{1,2}$}
\address{$^1$ Bogolyubov Institute for Theoretical Physics of the National Academy of Sciences of Ukraine \\
Metrologichna Str., 14-b,  Kyiv, 03143, Ukraine \\
$^2$ National Technical University of Ukraine
“Igor Sikorsky Kyiv Polytechnic Institute” \\
Peremohy av., 37, Kyiv, 03056,  Ukraine}

\begin{abstract}
{Solutions of the Dirac equation for an electron in the Coulomb potential are obtained using operator invariants of the equation, namely the Dirac, Johnson-Lippmann and recently found new invariant. It is demonstrated that these  operators are the spin invariants. The generalized invariant is constructed and the exact general solution of the Dirac equation are found. 
In particular, the explicit expressions of the bispinors corresponding to the three complete sets 
of the invariants, their eigenvalues and quantum numbers are calculated. It is  shown that
the general solution of one center Coulomb Dirac equation contains 
free parameters. Changing one or more of these parameters, one can transform one solution of the Dirac equation into any other.  It is shown for the first time that these invariants 
 determine electron spatial  probability amplitude and spin polarization in each quantum state. Electron probability densities 
and spin polarizations are explicitly calculated in the general form for several electron states 
in the hydrogen-like energy spectrum. Spatial distributions of these characteristics 
are shown to depend essentially on the invariant set, demonstrating, in spite of the accidental degeneracy of energy levels, physical difference of the states corresponding to different spin invariants.}

\end{abstract}

\maketitle   



\section{Introduction} 

The  well-known and comprehensively studied Dirac equation (DE) 
\begin{equation}
\label{wDeq} 
i\hbar \frac{\partial \Psi}{\partial t} = \hat{H} \Psi , \quad \hat{H} = c \left( \hat{\mathbf{p}} - \frac{e}{c} \mathbf{A}(\mathbf{r}) \right) \bm{\hat{\alpha}} + e\varphi (\mathbf{r}) \hat{I} + m c^{2} \hat{\beta} 
\end{equation}
is the main equation  of the relativistic theory of particles in external electromagnetic field. In Eq. (\ref{wDeq}) $ \varphi (\mathbf{r}) $ and $ \mathbf{A}(\mathbf{r}) $ are scalar and vector potentials of the field. Basic role of the DE was revealed soon after its discovery in 1928 \cite{Dirac1} (see \cite{Dirac,BetSol,LandauIV}). In the same year C. G. Darwin wrote "... Dirac has brilliantly removed the defects before existing in the mechanics of the electron, and has shown how the phenomena usually called the “ spinning electron ” fit into place in the complete theory" \cite{Darwin}. The study of particle motion in the Coulomb potential within the DE \cite{Darwin,Gordon} resulted in prediction of the fine-structure splitting of hydrogen-like spectrum due to spin–orbit interaction. Solutions of the DE were investigated in numerous papers \cite{Darwin,Gordon,Davis,Martin,Biedenharn,Drake,Al-Hashimi} whose results are summarized in textbooks (see, e.g., \cite{Dirac,BetSol,LandauIV,McConnell,Davydov,Messia,Bete}).

The DE, as was demonstrated by Darwin \cite{Darwin} and Gordon \cite{Gordon}, admits exact stationary solutions, which means that Dirac problem with the Coulomb potential describes the integrable system. It can be characterized by the set of independent commuting operators $ \mathcal{H} = \lbrace \hat{H}_{1}, \hat{H}_{2}, \ldots \, , \hat{H}_{n} \rbrace $, where  $ n $ corresponds to the number of the degrees of freedom, and where the set $ \mathcal{H} $ includes the Hamiltonian and operators of the integrals of motion, i.e., invariants. In quantum case the operators of observables in such sets have a joint system of eigenstates, and their eigenvalues determine quantum numbers of each state. It is well known that in the relativistic Coulomb problem the Hamiltonian and the angular momentum do not generate a complete set of commuting observables, and an additional operator is needed to complete the set. In particular, the integrability of the non-relativistic Schr\"{o}dinger equation is secured by a set of three independent commuting operators including the Hamiltonian, so that the quantum states acquire three quantum numbers. The state space of a particle with spin $ 1/2 $ is represented by the tensor product of the spatial, including coordinate and momentum components, and the two-dimensional spin states. In other words, integrability of the DE requires four such operators in the full set. This gives an additional quantum number which takes two values. Its meaning in the non-relativistic limit is the spin quantum number $ \sigma $.

For the central-symmetric field integrals of motion of the DE include the total angular momentum $ \mathbf{\hat{J}} $ and operator $ \hat{K}_{D} $ used by Dirac (Dirac invariant, DI). The angular momentum is given by the sum of the orbital $ \hat{\mathbf{L}} \hat{I} $ and intrinsic $ \bm{\hat{\Sigma}} $ angular momenta, $ \mathbf{\hat{J}} = \hat{\mathbf{L}} \hat{I} + (\hbar /2) \bm{\hat{\Sigma}} $, and $ \hat{I} $ is $ 4\times 4 $ unit matrix. This means that the set $ \mathcal{H}_{D} = \lbrace \hat{H}_{D}, \hat{J}^{2}, \hat{J}_{z}, \hat{K}_{D} \rbrace $ with $ \hat{H}_{D} $ being Dirac Hamiltonian, provides complete integrability of the DE. This permits to reduce the system of four partial differential equations for spinor amplitudes to the system of the first order ordinary differential equations for radial and angular functions \cite{Breev}. 

There is also the so called Johnson-Lippmann invariant (JLI), $ \hat{A}_{JL} $, constructed by Johnson and Lippmann (JLI) in 1950 \cite{John-Lip}. In classical and quantum physics it is established that  presence of integrals of motion additional to the number of the degrees of freedom, makes system degenerate. For instance, \textit{the accidental  degeneracy} takes place for a particle in the central field $ V \sim  1 /r $.

The degeneracy is substantially lowered for the relativistic Kepler problem. Nevertheless, the removal of the dynamic symmetry is not complete and most of the energy levels remain to be degenerate. The degeneracy is accounted for by means of the Johnson-Lippmann constant of motion \cite{John-Lip} whose operator does not commute with the DI and results in the accidental degeneracy. It allows to present equations for radial functions in various representations \cite{Martin,Biedenharn,Breev}, as well as to obtain a purely algebraic solution of the DE using the methods of supersymmetric quantum mechanics (see \cite{Sukumar,JarvSted,Dahl,TKhaAKhe,Bagchi}). 

The presence of two noncommutative invariants, resulting in energy degeneracy, allows to describe stationary states by different systems of Hamiltonian eigenfunctions corresponding to different invariants and, therefore, with different physical meaning. For example, in Ref. \cite{PhysRev13} the observation of the nodal structure of Stark hydrogen levels has been reported. Theoretical explanation of the obtained data was based on the solution of the Schr\"{o}dinger equation in parabolic coordinates, i.e. on the joint system of eigenfunctions of the set $ \mathcal{H}_{par} = \lbrace \hat{H}, \hat{A}_{z}, \hat{L}_{z} \rbrace $ where $ \hat{A}_{z} $ is the operator of $ z $-component of Laplace-Runge-Lenz vector. Direct comparison of the experimental and calculated quantities has revealed good agreement \cite{PhysRev13} and validated that hydrogen discrete states correspond to the set $ \mathcal{H}_{par} $, not to the set $ \mathcal{H}_{sph} = \lbrace \hat{H}, \hat{\mathbf{L}}^{2}, \hat{L}_{z} \rbrace $ which gives solution in spherical coordinates. In view of this one can  hope that in future precise experiments, depending on the chosen  experimental conditions, hydrogen atom quantum states corresponding to each set of invariants, can be measured.  

In the present paper we aim to calculate the solutions of the one center Coulomb Dirac equation using the complete set of invariants, which includes the Hamiltonian, square of the total angular momentum, its projection and one of the spin invariants, namely Dirac, Johnson-Lippmann or recently found new invariant \cite{BEL_arj}. We aim to prove that spatial distributions of the electron charge probability and spin expectations depend essentially on the invariant set, demonstrating, in spite of the accidental degeneracy of energy levels, physical difference of the states corresponding to different spin invariants.

The paper, which is the continuation of our previous paper \cite{BEL_arj}, is organized as follows. First, in Section 2, we discuss solution of the stationary DE with the Coulomb potential in a way  different from the standard textbooks and Ref. \cite{BEL_arj} in order to demonstrate  that the DE eigen bispinors are not completely determined without the four invariants. Then, in Section 3, we  analyze recently found  new invariant  of the DE \cite{BEL_arj}, called BEL invariant (BELI),  and study the eigenvalue problem for the sets of the independent commuting operators  $ \lbrace \hat{H}, \hat{J^{2}}, \hat{J}_{z}, \hat{{\cal I}}_{inv} \rbrace $ where $ \hat{{\cal I}}_{inv} $ is DI, JLI or BELI. Explicit expressions of the eigen bispinors for each set are calculated and a full set of quantum numbers of the stationary states is defined. In Section 4 we construct the generalized invariant and calculate its eigen bispinors. In Section 5 the main attention is devoted to electron probability densities and spin polarizations for some hydrogen-like states. Finally, in Section 6, the summary of the obtained results and their discussion are presented. 

\section{Invariants of the Dirac Equation and its solution} 

In time-independent Coulomb field of a positive charge $ Ze $ electron states $ \Psi (\mathbf{r},t) = \Psi (\mathbf{r}) \exp \left(-iEt/\hbar \right) $ are determined by the DE 
\begin{equation}
\label{sDE}
\hat{H} \Psi = E \Psi , \quad   \hat{H} = c \mathbf{\hat{p}} \bm{\hat{\alpha}} + V(r) \hat{I} + m c^{2} \hat{\beta} , \quad V(r) = - Ze^{2}/r ,
\end{equation}
in which $ E $ is the  energy, operator $ \hat{H} $ is the Dirac Hamiltonian, 
$$ \Psi (\mathbf{r},t) = \left( \psi_{1}(\mathbf{r},t) \: \psi_{2}(\mathbf{r},t) \: \psi_{3}(\mathbf{r},t) \: \psi_{4}(\mathbf{r},t) \right)^{T} $$ 
is a 4-component amplitude of the spinor field (4-spinor, or bispinor),  $ c $ is speed of light ,   $ \mathbf{\hat{p}} = -i\hbar \bm{\nabla} $ is momentum operator, $ \hat{\beta} $ and components $ \hat{\alpha}_{j} $ ($ j=x,y,z $) of the vector-matrix $ \bm{\hat{\alpha}} = \sum_{j} \mathbf{e}_{j} \hat{\alpha}_{j} $ together with the unit matrix  $ \hat{I} $ are $ 4\times 4 $ Hermitian Dirac matrices (DM). 

In fact, Eq. (\ref{sDE}) is a system of four equations for the components  $ \psi_{\nu} $ ($ \nu = 1,2,3,4 $), which acquire the obvious form after choosing DM. Below we use their standard form, represent $ 4 \times 4 $-dimensional matrices via $ 2 \times 2 $ matrices and write 
 the Hamiltonian in the form
\begin{equation}
\label{H_D-m} 
\hat{H}_{D} = \left( \begin{array}{cc}
\left( V(r)+mc^{2} \right) \hat{I}_{2} & c\bm{\hat{\sigma}} \mathbf{\hat{p}} \\ c\bm{\hat{\sigma}} \mathbf{\hat{p}} & \left( V(r)-mc^{2} \right) \hat{I}_{2}
\end{array}  \right) ,
\end{equation}
where $ \hat{I}_{2} $ is a unit $ 2\times 2 $ matrix, $ \hat{\sigma}_{j} $ ($ j=x,y,z $) are Pauli matrices,  $ \bm{\hat{\sigma}} = \sum_{j} \mathbf{e}_{j} \hat{\sigma}_{j} $, and $  \mathbf{e}_{j} $ are the corresponding orths. 

 The block form of DM allows to represent the bispinor as
\begin{equation}
\label{bispinor} 
\Psi \left( \mathbf{r} \right) = \left( \begin{array}{c} \psi^{(u)} \left( \mathbf{r} \right)  \\ \psi^{(d)} \left( \mathbf{r} \right) 
\end{array} \right) , \quad 
\psi^{(u)} = \left( \begin{array}{c} 
\psi^{(1)} \left( \mathbf{r} \right)  \\ \psi^{(2)} \left( \mathbf{r} \right) 
\end{array} \right) , \; \psi^{(d)} = \left( \begin{array}{c}
\psi^{(3)} \left( \mathbf{r} \right)  \\ \psi^{(4)} \left( \mathbf{r} \right) 
\end{array} \right) ,
\end{equation}
where $ \psi^{(u/d)} $ are its upper/lower spinors, respectively, with the components  $ \psi^{(\nu )} $ and $ \psi^{(\nu + 1)} $ ($ \nu =1 $ for the upper bispinor, and $ \nu = 3 $ for the lower one). 

In representation  (\ref{H_D-m}) and (\ref{bispinor}) the DE (\ref{sDE}) is transformed to 
\begin{equation}
\label{DsysEq_1} 
\begin{array}{c}
\left(E + Ze^{2}/r - mc^{2} \right) \psi^{(u)} - c\bm{\hat{\sigma}} \mathbf{\hat{p}} \psi^{(d)}  = 0  , \\
- c\bm{\hat{\sigma}} \mathbf{\hat{p}}\psi^{(u)} + \left(E + Ze^{2}/r + mc^{2} \right) \psi^{(d)}  = 0  .
\end{array}
\end{equation}
To obtain these equations in polar coordinates, one can use transformation of the kinetic energy operator $ \bm{\hat{\sigma}} \mathbf{\hat{p}} $ using a unit matrix
\begin{equation} 
\label{sigma_r} 
\hat{\sigma}_{r} = \bm{\hat{\sigma}} \mathbf{r} /r = \left( 
\begin{array}{cc}
\cos \vartheta & e^{-i\varphi} \sin \vartheta \\ e^{i\varphi} \sin \vartheta & -\cos \vartheta
\end{array} \right) , \quad \hat{\sigma}_{r}^{2} = \hat{I}_{2} .
\end{equation} 
and the identity 
$$ 
\hat{\sigma}_{r} \left( \bm{\hat{\sigma}} \cdot \mathbf{\hat{p}} \right) =  (1/r) \left( \bm{\hat{\sigma}}\cdot \mathbf{r} \right) \left( \bm{\hat{\sigma}} \cdot \mathbf{\hat{p}} \right) = \frac{1}{r} \left( \mathbf{r}\cdot \mathbf{\hat{p}} + i \bm{\hat{\sigma}} \cdot \mathbf{r} \times \mathbf{\hat{p}} \right) = \hat{p}_{r} + \frac{i}{r} \hat{\Lambda} ,
$$ 
where 
\begin{equation}
\label{p_r} 
\hat{p}_{r} = \frac{1}{2}  \left( \frac{\mathbf{r}}{r} \cdot \mathbf{\hat{p}} + \mathbf{\hat{p}} \cdot \frac{\mathbf{r}}{r} \right) = \frac{1}{r} \left( \mathbf{r} \cdot \hat{\mathbf{p}} - i\hbar \right) , \quad \hat{\Lambda} = \bm{\hat{\sigma}} \cdot \hat{\mathbf{L}} + \hbar \hat{I}_{2} , \quad \hat{\mathbf{L}} = \mathbf{r} \times \mathbf{\hat{p}} .
\end{equation}
This transformation is equivalent to transformations of DE using DM algebra \cite{McConnell,Martin}. 
In the above expressions  $ \hat{p}_{r} $ is Hermitian operator of the radial momentum (projection of operator $ \mathbf{\hat{p}} $ on the direction $ \mathbf{r} $) and $ \hat{\mathbf{L}} $ is the operator of the angular momentum. In spherical coordinates $ \hat{p}_{r} = -i\hbar \left( \frac{\partial}{\partial r} + \frac{1}{r} \right) $, matrix $ \hat{\Lambda} $ is 
\begin{equation}
\label{mLambda} 
\hat{\Lambda} = \hbar \left( \begin{array}{cc}
-i \frac{\partial}{\partial \varphi} + 1 &  e^{- i\varphi} \left( - \frac{\partial}{\partial \vartheta} + i \frac{\cos \vartheta }{\sin \vartheta} \frac{\partial}{\partial \varphi} \right) \\ 
 e^{i\varphi} \left( \frac{\partial}{\partial \vartheta} + i \frac{\cos \vartheta }{\sin \vartheta} \frac{\partial}{\partial \varphi} \right) & i \frac{\partial}{\partial \varphi} + 1   
\end{array} \right) ,
\end{equation} 
the kinetic energy  operator takes the form 
\[
c\bm{\hat{\sigma}} \mathbf{\hat{p}} = \hat{\sigma}_{r}^{2} c \left( \bm{\hat{\sigma}} \mathbf{\hat{p}} \right) = \hat{\sigma}_{r} c \left( \hat{p}_{r} + \frac{1}{r} \hat{\Lambda} \right) 
\]
and Eqs. (\ref{DsysEq_1}) can be re-written as 
\begin{equation}
\label{DsystEq2} 
\begin{array}{c}
\left(E + Ze^{2}/r - mc^{2} \right) \hat{\sigma}_{r} \psi^{(u)} - c\left( \hat{p}_{r} + \frac{i}{r} \hat{\Lambda} \right) \psi^{(d)}  = 0  , \\
- c\left( \hat{p}_{r} + \frac{i}{r} \hat{\Lambda} \right)\psi^{(u)} + \left(E + Ze^{2}/r + mc^{2} \right) \hat{\sigma}_{r} \psi^{(d)}  = 0  . 
\end{array}
\end{equation}
They include the radial momentum (\ref{p_r}) and matrices  (\ref{sigma_r}) and (\ref{mLambda})  which are anticommutative. Matrix $ \hat{\Lambda} $ depends on variables $ \vartheta $ and $ \varphi $ only, and anticommutation of $ \hat{\Lambda} $ and $ \hat{\sigma}_{r} $ implies that if $ \chi_{\lambda} $ is an eigenspinor of $ \hat{\Lambda} $ with an eigenvalue $ \hbar \lambda $, then the spinor $ \hat{\sigma}_{r} \chi_{\lambda} $ is also the eigenspinor of $ \hat{\Lambda} $ with the opposite sign  eigenvalue, i.e., if $ \hat{\Lambda} \chi_{\lambda} = \hbar \lambda \chi_{\lambda} $ then $ \hat{\Lambda} \hat{\sigma}_{r} \chi_{\lambda} = - \hbar \lambda \hat{\sigma}_{r} \chi_{\lambda} $ and $ \hat{\sigma}_{r} \chi_{\lambda} \sim \chi_{-\lambda} $. This allows to look for the solutions of Eqs. (\ref{DsystEq2}) in terms of the eigenspinors of matrix $\hat{\Lambda}$ with the separation of variable $ r $.

Equation $ \hat{\Lambda} \chi = \hbar \lambda \chi $ itself represents the system of equations for components of the spinor $ \chi = \left( f_{1} \left( \theta ,\varphi \right) \: f_{2} \left( \theta ,\varphi \right) \right)^{T} $:
\begin{equation}
\label{Eqs-f_1,2} 
\begin{array}{c}
-i \frac{\partial f_{1}}{\partial \varphi} + f_{1} + e^{-i\varphi} \left(-\frac{\partial f_{2}}{\partial \vartheta} + i \frac{\cos \vartheta}{\sin \vartheta} \frac{\partial f_{2}}{\partial \varphi} \right) = \lambda f_{1} ,  \\
e^{i\varphi} \left(\frac{\partial f_{1}}{\partial \vartheta} + i \frac{\cos \vartheta}{\sin \vartheta} \frac{\partial f_{1}}{\partial \varphi} \right) + i \frac{\partial f_{2}}{\partial \varphi} + f_{2} = \lambda f_{2} ,
\end{array}
\end{equation}
with the solutions given by functions $ f_{\nu} (\vartheta ,\varphi) = \exp \left(im_{\nu}\varphi \right)f_{\nu} \left(\vartheta \right) $ ($ \nu = 1,2 $), where $ m_{\nu} $ are connected  via the relation 
\begin{equation}
\label{m_1,m_2} 
 m_{2} - m_{1} = 1 . 
\end{equation}
Functions $ f_{\nu} \left(\vartheta \right) $ satisfy differential equations which can be reduced to one equation of the second order: 
\[
\frac{1}{\sin \vartheta} \frac{d}{d \vartheta} \left( \sin \vartheta \frac{d f_{\nu} \left( \vartheta \right)}{d \vartheta} \right) - \frac{m_{\nu}^{2}}{\sin^{2} \vartheta} f_{\nu} \left( \vartheta \right) + \lambda \left( \lambda - 1 \right) f_{\nu} \left( \vartheta \right) = 0 .
\]
So, spinor components are described by  equation known in theory of spherical functions. It admits solutions that satisfy finiteness and unambiguity conditions, provided  equality 
\begin{equation}
\label{det_lambda} 
\lambda \left( \lambda - 1 \right) = l \left( l + 1 \right) , 
\end{equation}
is valid, where $ l = 0,1,2,\ldots $ and $ m_{\nu} $ takes integer numbers in the interval $ -l \leq m_{\nu} \leq l $. The solution is given by the associated Legendre polynomial $  P_{l}^{ m_{\nu} } \left( \cos \vartheta \right) $. Relation (\ref{det_lambda}) gives two eigenvalues $ \lambda $,  
\begin{equation}
\label{eigvL} 
\lambda_{1} \equiv \lambda_{+} = l + 1  , \quad  \lambda_{2} \equiv \lambda_{-} = - l ,
\end{equation}
one of which is positive, and another one -- negative due to $ l $ positiveness.
 
Therefore, spinor components, with the accuracy of a constant multiplier, are 
\[
f_{1}\left( \vartheta , \varphi \right) = C_1 e^{im_{1}\varphi} P_{l}^{ m_{1} } \left( \cos \vartheta \right) ,  \quad f_{2}\left( \vartheta , \varphi \right) = C_2 e^{im_{2}\varphi} P_{l}^{ m_{2} } \left( \cos \vartheta \right) .
\]
Substituting these expressions into Eqs. (\ref{Eqs-f_1,2}) and requiring their identical validity, one gets the  relation for the constants $ C_1 = \left( \lambda + m_{1} \right) C_2 $ which depends on the eigenvalue $ \lambda_{\pm} $ and gives two solutions for spinors corresponding to the positive and negative eigenvalue,
\begin{equation}
\label{A_s,B_s} 
 C_{1+} = \left( \lambda_{+} + m_{1} \right) C_{2+} = \left(l+1+m_{1} \right)C_{2+} , \, C_{1-} =  \left( \lambda_{-} + m_{1} \right) C_{2-} = -\left(l - m_{1}\right)C_{2-},
\end{equation} 
respectively. Constants $ C_{2\pm} $ are defined by the normalization condition. 
Note, normalization for spinors determines the incoming constants up to the phase multiplier.

Therefore, the spinors are spherical ones,
\begin{equation}
\label{sphsp+2} 
\chi_{l, M,+} \left( \vartheta ,\varphi \right) = i^{l} \left( \begin{array}{c} 
\sqrt{\frac{\left( l + 1 + m_{1}\right) \left( l - \mid m_{1} \mid \right)! }{4\pi \left( l + \mid m_{1} \mid \right)!}} e^{im_{1}\varphi} P_{l}^{m_{1}} \left( \cos \vartheta \right)  \\ 
\sqrt{\frac{\left( l + 1 - m_{2}\right) \left( l - \mid m_{2} \mid \right)! }{4\pi \left( l + \mid m_{2} \mid \right)!}} e^{im_{2}\varphi} P_{l}^{m_{2}} \left( \cos \vartheta \right)     
\end{array} \right), 
\end{equation}
\begin{equation}
\label{sphsp-2} 
\chi_{l, M,-} \left( \vartheta ,\varphi \right) = i^{l} \left( \begin{array}{c} 
-\sqrt{\frac{\left( l - m_{1}\right) \left( l - \mid m_{1} \mid \right)! }{4\pi \left( l + \mid m_{1} \mid \right)!}} e^{im_{1}\varphi} P_{l}^{m_{1}} \left( \cos \vartheta \right)  \\ 
\sqrt{\frac{\left( l + m_{2}\right) \left( l - \mid m_{2} \mid \right)! }{4\pi \left( l + \mid m_{2} \mid \right)!}} e^{im_{2}\varphi} P_{l}^{m_{2}} \left( \cos \vartheta \right)     
\end{array} \right). 
\end{equation} 
Their components are characterized by integer numbers $ m_{\nu} $ which are connected via relations (\ref{m_1,m_2}). Two numbers with the given difference can be represented by $ m_{2} + m_{1} = 2M $ whence $ M $ which takes half-integer values, distinguishes spherical spinors and is included into their subscripts (\ref{sphsp+2})-(\ref{sphsp-2}), where $ m_{1} = M - 1/2 $ and $ m_{2} = M + 1/2 $. 

Spinors $ \chi_{l, M,\pm} $ satisfy equalities
\begin{equation}
\label{Eq-chi_+,-} 
\hat{\Lambda} \chi_{l, M,+} = \hbar \left( l + 1 \right)  \chi_{l, M,+} , \quad \hat{\Lambda} \chi_{l, M,-} = - \hbar l \chi_{l, M,-}.
\end{equation}
Using its exact forms (\ref{sphsp+2})-(\ref{sphsp-2}), one can define coefficient in the relation $ \hat{\sigma}_{r} \chi_{\lambda} \sim \chi_{-\lambda} $ and come to the following equalities \cite{LandauIV} 
\begin{equation}
\label{relat_chi} 
 \hat{\sigma}_{r} \chi_{ l,M,+} = i \chi_{l+1,M,-} , \quad \hat{\sigma}_{r} \chi_{ l,M,-} = -i \chi_{l-1,M,+} .
\end{equation}

Spherical harmonics determine angular dependence of spinors $ \psi^{(u/d)} $. Solution of Eqs. (\ref{DsystEq2}) can be represented as
\begin{equation}
\label{anzatsp} 
\begin{array}{c}
\psi^{(u)} = F^{(+)} \left( r \right) \chi_{l_{1}, M,+}\left( \vartheta ,\varphi \right) + F^{(-)} \left( r \right) \chi_{l_{2}, M,-}\left( \vartheta ,\varphi \right) , \\ 
\psi^{(d)} = G^{(+)} \left( r \right) \chi_{l_{1}, M,+}\left( \vartheta ,\varphi \right) + G^{(-)} \left( r \right) \chi_{l_{2}, M,-}\left( \vartheta ,\varphi \right) , 
\end{array} 
\end{equation}
where the coefficients at $ \chi_{l_{\nu}, M,\pm} $ in the upper $ F^{(\pm)} $ and lower $ G^{(\pm)} $ spinors can depend on $ r $, only, and their upper indeces $ (\pm) $ indicate positive or negative eigen value of $ \hat{\Lambda} $ of the corresponding spherical spinor.

Substitution of (\ref{anzatsp}) into Eqs. (\ref{DsystEq2}) with account of (\ref{Eq-chi_+,-}) and equalities (\ref{relat_chi}) demonstrates that solution requires fulfilment of the relations 
\[
 \hat{\sigma}_{r} \chi_{ l_{1},M,+} = i \chi_{l_{1}+1,M,-} = i \chi_{l_{2},M,-} , \quad \hat{\sigma}_{r} \chi_{ l_{2},M,-} = -i \chi_{l_{2}-1,M,+} = -i \chi_{l_{1},M,+} .
\]
So, positive integer numbers $ l_{1} $ and $ l_{2} $ in (\ref{anzatsp}) should be connected by the equality 
\begin{equation}
\label{l_1,l_2} 
 l_{2} - l_{1} = 1 ,
\end{equation}
and Eq. (\ref{DsystEq2}) gives the following equalities: 
\begin{equation}
\label{DEqs_red} 
\begin{array}{c}
\left[ -c\left( \hat{p}_{r} + i\frac{\hbar \left(l_{1}+1 \right) }{r} \right) G^{(+)} - i \left( E - mc^{2} + \frac{Ze^{2}}{r} \right) F^{(-)} \right] \chi_{l_{1},M,+} + \\
+ \left[ -c\left( \hat{p}_{r} - i\frac{\hbar l_{2}}{r} \right) G^{(-)} + i\left( E - mc^{2} + \frac{Ze^{2}}{r} \right) F^{(+)} \right] \chi_{l_{2}, M,-} = 0 , \\
\left[ -c\left( \hat{p}_{r} + i\frac{\hbar \left(l_{1}+1 \right)}{r} \right) F^{(+)} - i \left( E + mc^{2} + \frac{Ze^{2}}{r} \right) G^{(-)} \right] \chi_{l_{1}, M,+} + \\ 
+ \left[ -c\left( \hat{p}_{r} - i\frac{\hbar l_{2}}{r} \right) F^{(-)} + i \left( E + mc^{2} + \frac{Ze^{2}}{r} \right) G^{(+)} \right] \chi_{l_{2}, M,-} = 0 .
\end{array}
\end{equation}
Since the spinors $ \chi_{l, M,\pm} $ are independent, the relations (\ref{DEqs_red}) are valid at zero values of the "coefficients", which gives  four equations for functions 
$ F^{(\pm)}(r) $ and $ G^{(\pm)}(r) $. The pair of numbers $ l_{1} $ and $ l_{2} $ has the given difference (\ref{l_1,l_2}), and, thus, one can represent them as $ l_{1} = j - 1/2 $ and $ l_{2} = j + 1/2 $, where $ j \equiv ( l_{1} + l_{2} )/2$ takes positive half-integer values. So, the positive numbers $ l_{1}+1 $ and  $ l_{2} $  are $ l_{1}+1 = j + 1/2 $  and $ l_{2} = j + 1/2 $, and, hence, Eqs. (\ref{DEqs_red}) for radial functions involve the number 
\begin{equation}
\label{kappa_j} 
 \kappa_{j} = j + \frac{1}{2}  , \quad \kappa_{j} = 1,2,\ldots 
\end{equation}
which takes non-zero positive integer values.

System of equations for radial functions separates into two independent pairs, one -- for functions  $ F^{(+)}(r) $ and $ G^{(-)}(r) $, and the second one -- for functions $ F^{(-)}(r) $ and $ G^{(+)}(r) $. Similar equations for such functions are well known and their solutions are expounded in many textbooks (e.g.,  \cite{BetSol,LandauIV,McConnell}).

It is convenient to introduce dimensionless coordinate $ \xi = \left(mc /\hbar \right) r $ for functions $ rF^{(\pm)} $ and $ rG^{(\mp)} $, where $ \hbar /mc $ is electron Compton wavelength. Taking into account the boundary condition at infinity, solutions for the bound states can be searched for in the form 
\begin{equation}
\label{f,g} 
\begin{array}{c}
rF^{(+)} = \sqrt{1+\varepsilon} e^{-\varkappa \xi} \xi^{\gamma} \left[ u_{1} \left( \xi \right) + v_{1} \left( \xi \right) \right] , \quad 
rG^{-} = \sqrt{1-\varepsilon} e^{-\varkappa \xi} \xi^{\gamma} \left[ u_{1} \left( \xi \right) - v_{1} \left( \xi \right) \right] , \\
rF^{(-)} = \sqrt{1+\varepsilon} e^{-\varkappa \xi} \xi^{\gamma} \left[ u_{2} \left( \xi \right) + v_{2} \left( \xi \right) \right] , \quad 
rG^{+} = - \sqrt{1-\varepsilon} e^{-\varkappa \xi} \xi^{\gamma} \left[ u_{2} \left( \xi \right) - v_{2} \left( \xi \right) \right] ,
\end{array}
\end{equation}
where 
\begin{equation}
\label{varcappa} 
\varkappa = \sqrt{ 1 - \varepsilon^{2} }  ,  \quad  \varepsilon^{2} = E^{2}/m^{2}c^{4}  < 1 
\end{equation} 
are dimensionless damping $ \varkappa $ and energy $ \varepsilon$,  $ \varepsilon = E/mc^{2} $.  

Substitution of expressions (\ref{f,g}) into equations for radial functions following from Eqs. (\ref{DEqs_red}), leads to equations one of which gives functions  $ u_{1} $ and $ v_{1} $, and the second one --  $ u_{2} $ and $ v_{2} $:
\begin{equation}
\label{Eq-u,v_sigm} 
\begin{array}{c}
\xi \frac{d u_{\nu }}{d \xi} + \left( \gamma + \frac{Z\alpha \varepsilon}{\varkappa} - 2\varkappa \xi \right) u_{\nu } \mp \left( \kappa_{j} \mp \frac{Z\alpha }{\varkappa } \right) v_{\nu } = 0 , \\ 
\xi \frac{d v_{\nu }}{d \xi} + \left( \gamma - \frac{Z\alpha \varepsilon}{\varkappa} \right) v_{\nu } \mp \left( \kappa_{j} \pm \frac{Z\alpha }{\varkappa } \right) u_{\nu } = 0 ,\quad \nu =1,2,
\end{array}  
\end{equation} 
where the upper sign corresponds to functions with $\nu =1$, lower -- with $ \nu =2 $, and $ \alpha = e^{2} /\hbar c $  is {\it{Sommerfeld fine structure constant}}. This system can be reduced to the second order equations for each function:
\[
\begin{array}{c}
\xi^{2} \frac{d^{2} u_{\nu }}{d\xi^{2}} + \xi \left( 1 + 2\gamma - 2\varkappa \xi \right) \frac{du_{\nu }}{d\xi} + 2 \varkappa \xi \left( \frac{Z\alpha \varepsilon}{\varkappa} - \gamma -1 \right) u_{\nu } + \left( \gamma^{2} - \kappa_{j}^{2} + Z^{2}\alpha^{2} \right) u_{\nu } = 0 , \\
\xi^{2} \frac{d^{2} v_{\nu }}{d\xi^{2}} + \xi \left( 1 + 2\gamma - 2\varkappa \xi \right) \frac{dv_{\nu }}{d\xi} + 2\varkappa \xi \left( \frac{Z\alpha \varepsilon}{\varkappa} - \gamma \right) v_{\nu } + \left( \gamma^{2} - \kappa_{j}^{2} + Z^{2}\alpha^{2} \right) v_{\nu } = 0 , 
\end{array}
\]
from where it follows that functions $ u $ and $ v $ should satisfy hypergeometric differential equation. Finitness of the solution requires that hypergeometric series should terminate at some value of $ n $ which imposes  two conditions: $ \gamma^{2} - \kappa_{j}^{2} + Z^{2}\alpha^{2} = 0 $ and $ \gamma - Z\alpha \varepsilon /\varkappa = n $, where $ n $ is zero or positive integer. Thus, equations for $ u $ and $ v $ become the differential equation for generalized Laguerre polynomial $ \mathit{L}_{n}^{2\gamma} \left( \rho \right) $ for variable $ 2\varkappa \xi = \rho $. 
From the first condition we get 
\begin{equation}
\label{mu} 
\gamma = \gamma_{j} = \sqrt{\kappa^{2}_{j}  - Z^{2}\alpha^{2}} ,
\end{equation}
where the square root has positive sign only, as it follows from the condition of the convergence of the  normalization. The second one 
\begin{equation}
 \label{Eq_epsilon} 
 \frac{Z\alpha \varepsilon}{\varkappa} - \gamma _j = n_{r} , \quad n_{r} = 0,1,2,\ldots, 
\end{equation} 
is a well-known equation for the relativistic hydrogen-like energy spectrum. So the searched functions can be expressed via generalized Laguerre polynomials:
\[
 u_{1} =\mathcal{U}_{1} \mathit{L}_{n_{r}-1}^{2\gamma_{j}} \left( \rho \right)  , \quad v_{1} = \mathcal{V}_{1} \mathit{L}_{n_{r}}^{2\gamma_{j}} \left( \rho \right) , \quad u_{2} = \mathcal{U}_{2}  \mathit{L}_{n_{r}-1}^{2\gamma_{j}} \left( \rho \right)  , \quad v_{2} = \mathcal{V}_{2}  \mathit{L}_{n_{r}}^{2\gamma_{j}} \left( \rho \right) ,
\]
where the radial quantum number $ n_{r} $ determines the degree of the Laguerre polynomials.

Equation (\ref{Eq_epsilon}) gives hydrogen-like energy spectrum for bound states 
\begin{equation}
\label{E_rel2} 
\varepsilon \equiv \varepsilon_{n_{r},j} = \frac{n_{r} + \gamma_{j}}{{\mathcal N}_{n_{r},j}} = \sqrt{1 - \frac{Z^{2}\alpha^{2}}{{\mathcal N}^2_{n_{r},j}}} \equiv \sqrt{1 - \varkappa ^2 _{n_{r},j} } ,
\end{equation}
where $\varkappa_{n_{r},j}=Z\alpha /{\mathcal{ N}_{n_{r},j}}$ and the notation  
\begin{equation}
\label{K_n,j }
{\mathcal N}_{n_{r},j} = \sqrt{\left(n_{r} + \gamma_{j} \right)^{2} + Z^{2}\alpha^{2} }=
\sqrt{n_{r}^2 + \kappa^2_j+2n_{r} \gamma_j}
\end{equation} 
is used. 

From the functional relations for Laguerre polynomials the relations follow:
\[
\mathcal{U}_{1} = - \frac{n_{r} + 2\gamma_{j}}{{\mathcal N}_{n_{r},j} + \kappa_{j}} \mathcal{V}_{1} , \quad \mathcal{V}_{2} = - \frac{n_{r} }{{\mathcal N}_{n_{r},j} + \kappa_{j}} \mathcal{U}_{2} . 
\] 
Thus, the solution of DE (\ref{DEqs_red}) is given by spinors (\ref{anzatsp}) with radial functions  
\begin{equation}
\label{F^+,-;G^+,-} 
\begin{array}{c}
F_{n_{r},j}^{(+)} \left( \rho \right) = \frac{2Z}{r_{B}\mathcal{N}_{n_{r},j}} A e^{-\rho/2} \rho^{\gamma_{j}-1} \left[ \mathit{L}_{n_{r}}^{2\gamma_{j}} \left( \rho \right) - \frac{n_{r} + 2\gamma_{j}}{\mathcal{N}_{n_{r},j}+\kappa_{j}} \mathit{L}_{n_{r}-1}^{2\gamma_{j}} \left( \rho \right) \right]  , \\
F_{n_{r},j}^{(-)} \left( \rho \right) = \frac{2Z}{r_{B}\mathcal{N}_{n_{r},j}} B e^{-\rho/2} \rho^{\gamma_{j}-1} \left[  \frac{n_{r} }{\mathcal{N}_{n_{r},j}+\kappa_{j}} \mathit{L}_{n_{r}}^{2\gamma_{j}} \left( \rho \right)  -  \mathit{L}_{n_{r}-1}^{2\gamma_{j}} \left( \rho \right) \right]  , \\ 
G_{n_{r},j}^{(+)}\left( \rho \right) = \frac{2Z}{r_{B}\mathcal{N}_{n_{r},j}} \sqrt{\frac{1 - \varepsilon_{n_{r},j}}{1 + \varepsilon_{n_{r},j}}} B e^{-\rho/2} \rho^{\gamma_{j}-1}  \left[ \frac{n_{r} }{\mathcal{N}_{n_{r},j}+\kappa_{j}} \mathit{L}_{n_{r}}^{2\gamma_{j}} \left( \rho \right) + \mathit{L}_{n_{r}-1}^{2\gamma_{j}} \left( \rho \right) \right]  , \\
G_{n_{r},j}^{(-)}\left( \rho \right) = - \frac{2Z}{r_{B}\mathcal{N}_{n_{r},j}} \sqrt{\frac{1 - \varepsilon_{n_{r},j}}{1 + \varepsilon_{n_{r},j}}} A e^{-\rho/2} \rho^{\gamma_{j}-1} \left[  \mathit{L}_{n_{r}}^{2\gamma_{j}} \left( \rho \right) + \frac{n_{r} + 2\gamma_{j}}{\mathcal{N}_{n_{r},j}+\kappa_{j}} \mathit{L}_{n_{r}-1}^{2\gamma_{j}} \left( \rho \right) \right]  , 
\end{array} 
\end{equation}
where 
\begin{equation}
\label{AB}
 A = ( 2\varkappa )^{-\gamma} \sqrt{1+\varepsilon} \mathcal{V}_{1} , \qquad B = - ( 2\varkappa )^{-\gamma} \sqrt{1+\varepsilon} \mathcal{U}_{2}. 
 \end{equation}
Here the notation for dimensionless radial variable
\begin{equation}
\label{x_n,j} 
\rho \equiv \rho_{n_{r},j}  = 2\varkappa_{n_{r},j} \xi = \frac{2Zr}{r_{B}\mathcal{N}_{n_{r},j}} 
\end{equation} 
is used, in which $ r_{B} = \hbar^{2} /me^{2} $ is \textit{Bohr radius}. 

Constants $ A $ and $ B $  (\ref{AB}) can be found from the general normalization  
\[
\int \Psi_{n_{r},j,M}^{\dagger} \left( \mathbf{r} \right) \Psi_{n_{r},j,M} \left( \mathbf{r} \right) d\mathbf{r} = 1 .
\]
Introducing the common normalization constant $ \sqrt{2Z/r_{B} \mathcal{N}_{n_{r},j}} C_{n_{r},j} $, where 
\begin{equation}
\label{cnrj} 
 C_{n_{r},j} =  \sqrt{\frac{\left( 1 + \varepsilon_{n_{r},j} \right) \left( \mathcal{N}_{n_{r},j} + \kappa_{j} \right) n_{r}! }{4\mathcal{N}_{n_{r},j} \Gamma \left( n_{r} + 1 + 2\gamma_{j} \right)}}, 
\end{equation}
one can represent $ A $ and $ B $  as 
\begin{equation}
\label{C_1,C_2} 
A = C_{n_{r},j} \beta_{1} , \quad B = \sqrt{\frac{n_{r} + 2\gamma_{j}}{n_{r} }} C_{n_{r},j} \beta_{2},
\end{equation}
where $ \beta_{1} $ and $ \beta_{2} $  comply the relation 
\begin{equation}
\label{norm2} 
\mid \beta _1 \mid^2 + \mid \beta _2\mid^2 = 1 . 
\end{equation} 

Therefore,  the solution of the DE (\ref{sDE}) is given by the bispinor (\ref{bispinor}), $ \Psi = \Psi_{n_{r},j,M} \left( \mathbf{r} \right) $ with its upper and lower spinors defined by expressions (\ref{anzatsp}) with spherical harmonics (\ref{sphsp+2})-(\ref{sphsp-2}) and functions  (\ref{F^+,-;G^+,-}), whose coefficients are (\ref{C_1,C_2}). It is described by numbers $ M $, $ j $ and $ n_{r} $: 
\begin{equation}
\label{Bispin_gen} 
\Psi_{n_{r}, j, M } = \left( \frac{2Z}{r_{B}\mathcal{N}_{n_{r},j}} \right)^{3/2} 
\left( \begin{array}{c} 
 \beta_{1} R_{n_{r},j}^{(+)} \chi_{j-1/2, M,+} + \beta_2  R_{n_{r},j}^{(-)} \chi_{j+1/2, M,-}  \\ 
\sqrt{\frac{1 - \varepsilon_{n_{r},j}}{1 + \varepsilon_{n_{r},j}}} \left( \beta_2  Q_{n_{r},j}^{(+)} \chi_{j-1/2, M,+} - \beta_1 Q_{n_{r},j}^{(-)} \chi_{j+1/2, M,-} \right)      
\end{array} \right) .
\end{equation} 
Here spinors $ \chi_{j\mp1/2, M,\pm} $ are given in Eqs.(\ref{sphsp+2})-(\ref{sphsp-2}) and radial functions are 
\begin{equation}
\label{tildP,Q} 
\begin{array}{c}
R_{n_{r},j}^{(+)} \left( \rho \right) = C_{n_{r},j} e^{-\rho /2} \rho^{\gamma_{j}-1}  \left[ \mathit{L}_{n_{r}}^{2\gamma_{j}} \left( \rho \right) - \frac{n_{r} + 2\gamma_{j}}{\mathcal{N}_{n_{r},j}+\kappa_{j}} \mathit{L}_{n_{r}-1}^{2\gamma_{j}} \left( \rho \right) \right] , \\ 
R_{n_{r},j}^{(-)} \left( \rho \right) = C_{n_{r},j} e^{-\rho /2} \rho^{\gamma_{j}-1} \left[ \frac{\sqrt{n_{r}\left( n_{r} + 2\gamma_{j} \right) } }{\mathcal{N}_{n_r,j}+\kappa_{j}} \mathit{L}_{n_{r}}^{2\gamma_{j}} \left( \rho \right)  - \sqrt{\frac{n_{r} + 2\gamma_{j}}{n_{r}}} \mathit{L}_{n_{r}-1}^{2\gamma_{j}} \left( \rho \right) \right]  , \\ 
Q_{n_{r},j}^{(-)}\left( \rho \right) = C_{n_{r},j} e^{-\rho /2} \rho^{\gamma_{j}-1} \left[ \mathit{L}_{n_{r}}^{2\gamma_{j}} \left( \rho \right) + \frac{n_{r} + 2\gamma_{j}}{\mathcal{N}_{n_{r},j}+\kappa_{j}} \mathit{L}_{n_{r}-1}^{2\gamma_{j}} \left( \rho \right) \right] , \\
Q_{n_{r},j}^{(+)}\left( \rho \right) = C_{n_{r},j} e^{-\rho /2} \rho^{\gamma_{j}-1} \left[ \frac{\sqrt{n_{r}\left( n_{r} + 2\gamma_{j} \right) } }{\mathcal{N}_{n_r,j}+\kappa_{j}} \mathit{L}_{n_{r}}^{2\gamma_{j}} \left( \rho \right) + \sqrt{\frac{n_{r} + 2\gamma_{j}}{n_{r}}} \mathit{L}_{n_{r}-1}^{2\gamma_{j}} \left( \rho \right) \right] 
\end{array}
\end{equation}
with the constant defined in Eq. (\ref{cnrj}). Bispinor (\ref{Bispin_gen}) provides the general solution for the bound states of the DE (\ref{E_rel2}). This solution is however not completely determined due to arbitrary parameters $ \beta_{1} $ and $ \beta_{2} $ with condition (\ref{norm2}), which leaves the ambiguity in its choice.

\section{Eigenvalues of the invariants} 

As it was mentioned above, stationary states are joint eigenstate vectors of a complete set of independent commuting operators which includes the Hamiltonian and invariant operators. For one center Coulomb DE the invariants are given by the square of the total angular momentum, $ \mathbf{\hat{J}} = \hat{\mathbf{L}} \hat{I} + \frac{\hbar}{2}  \bm{\hat{\Sigma}} $, its $ z $-component, and DI, $ \hat{K}_{D} $, or JLI, $ \hat{A}_{JL} $. The set of quantum numbers which characterize stationary states, reflects the eigenvalues of all commuting operators in the set $ \lbrace \hat{H}, \hat{J^{2}}, \hat{J}_{z}, \hat{{\cal I}}_{inv} \rbrace $, where $ \hat{{\cal I}}_{inv} $ is DI, JLI or BELI. The form of the latter is introduced below. So, the set of quantum numbers is determined by the following eigenvalue equations
\begin{equation}
\label{state}  
\begin{array}{c}
\hat{H}\vert \varepsilon ,j, M,\epsilon_{inv} \rangle = mc^{2} \varepsilon \vert \varepsilon ,j,M,\epsilon_{inv} \rangle , \\ 
\hat{J^{2}} \vert \varepsilon ,j,M,\epsilon_{inv} \rangle = \hbar^{2} j\left(j + 1 \right) \vert \varepsilon ,j,M,\epsilon_{inv} \rangle , \\
\hat{J}_{z} \vert \varepsilon ,j,M,\epsilon_{inv} \rangle = \hbar m_{j} \vert \varepsilon ,j,M,\epsilon_{inv} \rangle , \\
\hat{{\cal I}}_{inv} \vert \varepsilon ,j,M,\epsilon_{inv} \rangle = \epsilon_{inv} \vert \varepsilon ,j,M,\epsilon_{inv} \rangle .
\end{array}
\end{equation}
Here notation $ \varepsilon = E/mc^{2} $ is used, $ \hbar^{2} j \left( j+1 \right) $ and $ \hbar m_{j} $ are the eigenvalues of the operators $ \hat{J^{2}} $ and $ \hat{J}_{z} $ respectively, defined algebraically from the commutation relations for the components of the operator  $ \mathbf{\hat{J}} $, and $ \epsilon_{inv} $ is the eigenvalue of the operator $ \hat{{\cal I}}_{inv} $.

The first equation in the set (\ref{state}) determines the energy spectrum (\ref{E_rel2}) and gives the radial quantum number $ n_{r} $.  Numbers $ j $ and $ M $, whose sense is clear and generally known, in our calculations are not connected with the invariants. They appear in the solution due to two pairs of auxiliary numbers $ m_{1},\,m_{2} $ and $ l_{1},\,l_{2} $ with the given differences (\ref{m_1,m_2}) and (\ref{l_1,l_2}). 

Taking into account the block-diagonal form of the matrices of the operators $ \hat{J^{2}} $ and $ \hat{J}_{z} $ 
\begin{equation}
\label{J_z} 
\hat{J}_{z} = \hat{L}_{z} \hat{I} + \frac{\hbar}{2} \hat{\Sigma}_{z} = \left( \begin{array}{cc}
\hat{L}_{z} \hat{I}_{2} + \frac{\hbar}{2} \hat{\sigma}_{z} & 0 \\ 0 & \hat{L}_{z} \hat{I}_{2} + \frac{\hbar}{2} \hat{\sigma}_{z},
\end{array}  \right) 
\end{equation}
and   
\begin{equation}
\label{J^2} 
\hat{\mathbf{J}}^{2} = \left( \begin{array}{cc}
\left( \mathbf{\hat{L}}^{2} + \frac{3}{4} \hbar^{2} \right) \hat{I}_{2} + \hbar \bm{\hat{\sigma}} \mathbf{\hat{L}} & 0 \\ 0 & \left( \mathbf{\hat{L}}^{2} + \frac{3}{4} \hbar^{2} \right) \hat{I}_{2} + \hbar \bm{\hat{\sigma}} \mathbf{\hat{L}}
\end{array}  \right) 
\end{equation} 
with equal blocks, we see  from the second and third equations in (\ref{state}) for the bispinor (\ref{Bispin_gen}) that the upper $ \psi^{(u)} $ and lower $ \psi^{(d)} $ spinors in the eigenvalue problems $ \hat{J}_{z} \Psi = \hbar M \Psi $ and $ \hat{\mathbf{J}}^{2} \Psi = \hbar^{2} j(j+1) \Psi $ satisfy the same equations. The identity $  \bm{\hat{\sigma}} \cdot \mathbf{\hat{L}}  \bm{\hat{\sigma}} \cdot \mathbf{\hat{L}} = \mathbf{\hat{L}}^{2} - \hbar \bm{\hat{\sigma}} \cdot \mathbf{\hat{L}} $ is valid, hence, the Legendre operator can be represented in the spinor form \cite{Biedenharn} and we get $ \mathbf{\hat{L}}^{2} + \hbar \bm{\hat{\sigma}} \mathbf{\hat{L}} + \frac{3}{4} \hbar^{2} = \hat{\Lambda}^{2} - \hbar^{2}/4 $ in operator $ \hat{J^{2}} $. Therefore, spinors $ \psi^{(u/d)} $ should satisfy equations 
\[
\left( -i\hbar \frac{\partial}{\partial \varphi} + \frac{\hbar}{2} \hat{\sigma}_{z} \right) \psi^{(u/d)}_{n_{r}, j, M } = \hbar m_{j} \psi^{(u/d)}_{n_{r}, j, M } , \quad \left( \hat{\Lambda}^{2} - \frac{\hbar^{2}}{4} \right) \psi^{(u/d)}_{n_{r}, j, M } = \hbar^{2} j\left(j + 1 \right) \psi^{(u/d)}_{n_{r}, j, M } .
\]

Angular dependence of spinors is included in spinors $ \chi_{j\mp1/2, M,\pm} $, only, as it follows from Eqs. (\ref{sphsp+2})-(\ref{sphsp-2}). The dependence of spinor components on spatial angle $ \varphi $ in the first equation gives $ M = m_{j} $. Since $ \chi_{j\mp1/2, M,\pm} $ are eigenspinors of matrix $ \hat{\Lambda} $ which, according to Eqs. (\ref{Eq-chi_+,-}), satisfy equalities 
\begin{equation}
\label{eigenspEq} 
\hat{\Lambda} \chi_{j- 1/2, M,+} = \hbar \kappa_{j} \chi_{j- 1/2, M,+} , \quad \hat{\Lambda} \chi_{j+ 1/2, M,-} = - \hbar \kappa_{j} \chi_{j+ 1/2, M,-} ,
\end{equation}
where $ \kappa_{j} $ is defined in (\ref{kappa_j}). 

The physical meaning of numbers $ j $ and $ M \equiv m_{j} $ follows from the equations
\[ 
\hat{J^{2}} \Psi_{n_{r}, j, M } = \hbar^{2} j \left( j+1 \right) \Psi_{n_{r}, j, M } , \qquad 
\hat{J}_{z} \Psi_{n_{r}, j, M } = \hbar M \Psi_{n_{r}, j, M }
\]
which define them as the definite values of the total angular momentum and its projection on the polar axis in a given stationary state. The latter,  of course, is a common knowledge, but here it was obtained in an original way.

(i) In relativistic problem, in addition to operators (\ref{J_z}) and (\ref{J^2}), there is DI  \cite{Dirac}
\begin{equation}
\label{inv_D} 
\hat{K}_{D} = \hat{\beta} \left( \bm{\hat{\Sigma}} \cdot \hat{\mathbf{L}} + \hbar \right) =  \left( \begin{array}{cc}
\hat{\Lambda} & 0 \\ 0 & - \hat{\Lambda} 
\end{array}  \right). 
\end{equation}
It commutes with the invariants $ \hat{\mathbf{J}}^{2} $ and $ \hat{J}_{z} $, has a block-diagonal form and acts on angular variables only. Therefore, the fourth  equation from the set (\ref{state}), $ \hat{K}_{D} \Psi_{n_{r}, j, M } = \epsilon_{D} \Psi_{n_{r}, j, M } $, can be written down as two simple spinor equations
\[
\hat{\Lambda} \psi^{(u)}_{n_{r}, j, M } = \epsilon_{D} \psi^{(u)}_{n_{r}, j, M } , \quad  - \hat{\Lambda} \psi^{(d)}_{n_{r}, j, M } = \epsilon_{D} \psi^{(d)}_{n_{r}, j, M } ,
\] 
which show that spinors $ \psi^{(u/d)} $ should be proportional to the eigen spinors of $ \hat{\Lambda} $ but with the opposite sign of the eigenvalue. According to Eqs. (\ref{eigenspEq}), for $ \psi^{(u)}_{n_{r}, j, M } = F^{(+)} \left( r \right) \chi_{j-1/2, M,+} $ and $ \psi^{(d)}_{n_{r}, j, M } = G^{(-)} \left( r \right) \chi_{j+1/2, M,-} $ ($ \beta_{1} = 1, \: \beta_{2} = 0 $)  the  eigenvalue of $ \hat{K}_{D} $, $ \epsilon_{D} = \hbar \kappa_{j} $ is positive, and \textit{vice versa}, for $ \psi^{(u)}_{n_{r}, j, M } = F^{(-)} \left( r \right) \chi_{j+1/2, M,-} $ and $ \psi^{(d)}_{n_{r}, j, M } = G^{(+)} \left( r \right) \chi_{j-1/2, M,+} $ ($ \beta_{1} = 0, \: \beta_{2} = 1 $),  the eigenvalue $ \epsilon_{D} = - \hbar \kappa_{j} $ is negative. Thus, the last eigenvalue equation in Eqs. (\ref{state}) fixes free parameters in the general expression  (\ref{Bispin_gen}) and results in two orthogonal normalized eigenbispinors with completely determined components. One can assign to the bispinors the number $ \sigma = \pm $ which indicates the sign of the eigenvalue $ \hat{K}_{D} $, and write them down in the form
\begin{equation}
\label{bispin^D} 
\begin{array}{c} 
\Psi_{n_{r}, j, m_{j},+ }^{(D)} (\mathbf{r}) = \left( \frac{2Z}{r_{B}\mathcal{N}_{n,j}} \right)^{3/2}  
\left( \begin{array}{c} 
 R_{n_{r},j}^{(+)} \chi_{j-1/2, m_{j},+}  \\ 
-\sqrt{\frac{1 - \varepsilon_{n_{r},j}}{1 + \varepsilon_{n_{r},j}}} Q_{n_{r},j}^{(-)} \chi_{j+1/2, m_{j},-}    
\end{array} \right) , \\
\Psi_{n_{r}, j, m_{j},- }^{(D)} (\mathbf{r}) = \left( \frac{2Z}{r_{B}\mathcal{N}_{n,j}} \right)^{3/2} 
\left( \begin{array}{c} 
R_{n_{r},j}^{(-)} \chi_{j+1/2, m_{j},-}  \\ 
\sqrt{\frac{1 - \varepsilon_{n_{r},j}}{1 + \varepsilon_{n_{r},j}}} Q_{n_{r},j}^{(+)} \chi_{j-1/2, m_{j},+}    
\end{array} \right) .
\end{array}
\end{equation}

Bispinors (\ref{bispin^D}) are the Darwin-Gordon solutions \cite{Darwin,Gordon} and represent the joint eigenbispinor system of the set $ \lbrace \hat{H}_{D}, \hat{J}_{z}, \hat{J^{2}}, \hat{K}_{D} \rbrace $ for which two independent but fixed pairs of the numbers $  \beta_{1}^{(\sigma)},  \beta_{2}^{(\sigma)} $ in bispinor (\ref{Bispin_gen}) are 
\begin{equation}
\label{beta_D} 
\left(\begin{array}{c}
\beta_{1}^{(+)} \\
\beta_{2}^{(+)}
\end{array} \right) = \left(\begin{array}{c}
1 \\
0
\end{array} \right)  , \quad \left(\begin{array}{c}
\beta_{1}^{(-)} \\
\beta_{2}^{(-)}
\end{array} \right) = \left(\begin{array}{c}
0 \\
1
\end{array} \right) .
\end{equation}
The assigned bispinor number $ \sigma = \pm $ is the fundamental characteristic of stationary states and is the fourth quantum number. It was also presented in a recent paper \cite{Breev} within symmetry algebra, in particular, using Yano-Killing operators.

(ii) In the Coulomb potential the JLI \cite{John-Lip} in conventional representation of Dirac matrices has the form
\begin{equation}
\label{inv_J-L} 
\hat{A}_{JL} = \frac{mZe^{2}}{r} \bm{\hat{\Sigma}} \cdot \hat{\mathbf{r}} - \frac{1}{c} \hat{K}_{D} \hat{\alpha}_{x} \hat{\alpha}_{y} \hat{\alpha}_{z} \left( \hat{H} - mc^{2} \hat{\beta} \right) = \left( \begin{array}{cc}
\bm{\hat{\sigma}} \cdot \hat{\mathbf{A}}_{+} & i(Ze^{2}/cr) \hat{\Lambda} \\ -i(Ze^{2}/cr) \hat{\Lambda} & -\bm{\hat{\sigma}} \cdot \hat{\mathbf{A}}_{-} 
\end{array}  \right) \,  ,
\end{equation}
where
\begin{equation}
\label{opPi} 
\hat{\mathbf{A}}_{\pm} = \frac{1}{2} \left( \hat{\mathbf{L}} \times \hat{\mathbf{p}} - \hat{\mathbf{p}} \times \hat{\mathbf{L}} \right) \pm mZe^{2} \frac{\mathbf{r}}{r} 
\end{equation}
are the Laplace-Runge-Lenz vector operators. It is easy to see that JLI commutes with $ \hat{J}_{z} $ and $ \hat{J^{2}} $ and does not commute with $ \hat{K}_{D} $. Therefore, the eigenstates of the Dirac Hamiltonian can be described by eigenbispinors of the set $ \mathcal{H}_{JL} = \lbrace \hat{H}, \hat{J}_{z}, \hat{J^{2}}, \hat{A}_{JL} \rbrace $ which differ from eigenbispinors (\ref{bispin^D}). 

In this case the fourth eigenvalue equation in Eqs. (\ref{state}) is $ \hat{A}_{JL} \Psi_{n_{r}, j, m_{j} } = \epsilon_{JL} \Psi_{n_{r}, j, m_{j} } $ which can be written down as the system of two spinor equations. By calculating vector operators  (\ref{opPi}) explicitly and writing them in spherical coordinates, one can obtain that
\[
\bm{\hat{\sigma}} \cdot \hat{\mathbf{A}}_{\pm} = \left( -i \hat{p}_{r} \hat{\Lambda} - \frac{1}{r} \hat{\Lambda}^{2} \pm mZe^{2} \right) \hat{\sigma}_{r} ,
\] 
where operator $ \hat{p}_{r} $ and matrices $ \hat{\Lambda} $ and $ \hat{\sigma}_{r} $ were defined in (\ref{p_r}), (\ref{mLambda}), and (\ref{sigma_r}), respectively. This leads to the equations 
\[
\begin{array}{c}
\left( -i \hat{p}_{r} \hat{\Lambda} - \frac{1}{r} \hat{\Lambda}^{2} + mZe^{2} \right) \hat{\sigma}_{r} \psi_{n_{r},j, m_{j}}^{(u)} + i\frac{Ze^{2}}{cr} \hat{\Lambda} \psi_{j, m_{j}}^{(d)} = \epsilon_{JL} \psi_{n_{r},j, m_{j}}^{(u)} , \\
-\left( -i \hat{p}_{r} \hat{\Lambda} - \frac{1}{r} \hat{\Lambda}^{2} - mZe^{2} \right) \hat{\sigma}_{r} \psi_{n_{r},j, m_{j}}^{(d)} - i\frac{Ze^{2}}{cr} \hat{\Lambda} \psi_{n_{r},j, m_{j}}^{(u)} = \epsilon_{JL} \psi_{n_{r},j, m_{j}}^{(d)} .
\end{array}
\]

Substituting spinors (\ref{anzatsp}) and taking into  account functional relations for Laguerre polynomials, we conclude that the bispinor (\ref{Bispin_gen}) becomes eigen one for JLI if free constants $ A $ and $ B $ in functions (\ref{F^+,-;G^+,-}) satisfy the relations
\begin{equation}
\label{C_1,2_JL} 
\begin{array}{c}
\left( n_{r} + 2\gamma_{j} \right) A + i  \frac{\epsilon_{JL} \mathcal{N}_{n_{r},j}}{Z me^{2} } B = 0 ,\\
i  \frac{\epsilon_{JL} \mathcal{N}_{n_{r},j}}{Z me^{2} } A - n_{r} B = 0 .
\end{array}
\end{equation}
The condition for their non-trivial solution 
gives two eigenvalues of the JLI
\begin{equation}
\label{eig_JL} 
\epsilon_{JL} = \pm a_{n_{r},j} , \quad a_{n_{r},j} = Z me^{2} \frac{\sqrt{n_{r} \left( n_{r} + 2\gamma_{j} \right)}}{\mathcal{N}_{n_{r},j}} = Z me^{2} \sqrt{1 - \frac{\kappa_{j}^{2}}{\mathcal{N}_{n_{r},j}^{2}}} , 
\end{equation}
and, respectively, two solutions for constants $ A $ and $ B $, and, threrefore (see (\ref{C_1,C_2})), two solutions for coefficients $ \left( \beta_{1}, \: \beta_{2} \right) $
\begin{equation}
\label{beta_JL} 
\left(\begin{array}{c}
\beta_{1}^{(+)} \\
\beta_{2}^{(+)}
\end{array} \right) = \left(\begin{array}{c}
e^{i\pi/4}/\sqrt{2} \\
-  e^{-i\pi/4}/\sqrt{2}
\end{array} \right)  , \quad \left(\begin{array}{c}
\beta_{1}^{(-)} \\
\beta_{2}^{(-)}
\end{array} \right) = \left(\begin{array}{c}
e^{i\pi/4} /\sqrt{2} \\
e^{-i\pi/4}/\sqrt{2}
\end{array} \right) ,
\end{equation}
first of which corresponds to the states with positive eigenvalue of the JLI, $ \epsilon_{JL} = a_{n_{r},j} $, and the second one -- with negative, $ \epsilon_{JL} = - a_{n_{r},j} $.
 
As a result, for the set $ \mathcal{H}_{JL} = \lbrace \hat{H}, \hat{J}_{z}, \hat{J^{2}}, \hat{{A}_{JL}}\rbrace $ we obtain the pair of orthonormal bispinors with the numbers $ \sigma = \pm $ assigned to them according to the sign in the eigenvalue equation  
 $ \hat{A}_{JL} \Psi_{n_{r}, j, m_{j},\sigma } = \sigma a_{n_{r},j} \Psi_{n_{r}, j, m_{j},\sigma } $, where $ \sigma = \pm $ characterizes the stationary states. These states are different from the states for the DE with DI, although the spectra of both sets coincide. 

(iii) In addition to the invariants (\ref{inv_D}) and (\ref{inv_J-L}), we constructed in \cite{BEL_arj} the operator of a new invariant, BELI, 
\begin{equation}
\label{A_BEL} 
\hat{{\cal I}}_{BEL}=\frac{1}{2i}\left[ \hat{K}_D, \hat{A}_{JL} \right] 
\end{equation}
which does commute with $ \hat{H} $, but does not commute with either operator it is constructed of. Hence, the set $ \mathcal{H}_{BEL} = \lbrace \hat{H}, \hat{J}_{z}, \hat{J^{2}}, \hat{{\cal I}}_{BEL} \rbrace $ defines one more system of eigenbispinors, different from the previous two cases. 

The explicit expression for BELI can be obtained by direct calculation, which gives
\begin{equation}
\label{invB} 
\hat{{\cal I}}_{BEL} = \left( \begin{array}{cc}
 \bm{\hat{\sigma}} \cdot \hat{\mathbf{B}}_{+} & \frac{Ze^{2}}{cr} \hat{\Lambda}^{2} \\ \frac{Ze^{2}}{cr} \hat{\Lambda}^{2} & \bm{\hat{\sigma}} \cdot \hat{\mathbf{B}}_{-}
\end{array}  \right) , 
\end{equation} 
where $ \hat{\Lambda} $ has been defined in Eq. (\ref{mLambda}), and the notation for the operator 
\begin{equation}
\label{vecB} 
\hat{\mathbf{B}}_{\pm} = \frac{1}{2} \left( \hat{\mathbf{L}} \times \hat{\mathbf{A}}_{\pm} - \hat{\mathbf{A}}_{\pm} \times \hat{\mathbf{L}} \right)  
\end{equation} 
is used (cp. Eq. (\ref{opPi})). In polar coordinates the scalar product $ \bm{\hat{\sigma}} \cdot \hat{\mathbf{B}}_{\pm} $ in the invariant (\ref{A_BEL}) has the form
\[
\bm{\hat{\sigma}} \cdot \hat{\mathbf{B}}_{\pm} =  \left( -\hat{p}_{r} \hat{\Lambda}^{2} + \frac{i}{r} \hat{\Lambda}^{3} \mp imZe^{2} \hat{\Lambda} \right) \hat{\sigma}_{r} .
\]
The last eigenvalue equation in Eqs. (\ref{state}), $ \hat{{\cal I}}_{BEL} \Psi_{n_{r}, j, m_{j} } = \epsilon_{BEL} \Psi_{n_{r}, j, m_{j} } $, leads to spinor equations 
\[
\begin{array}{c}
\left( -\hat{p}_{r} \hat{\Lambda}^{2} + \frac{i}{r} \hat{\Lambda}^{3} - imZe^{2} \hat{\Lambda} \right) \hat{\sigma}_{r} \psi_{n_{r},j, m_{j}}^{(u)} + \frac{Ze^{2}}{cr} \hat{\Lambda}^{2} \psi_{n_{r},j, m_{j}}^{(d)} = \epsilon \psi_{n_{r},j, m_{j}}^{(u)} , \\
\frac{Ze^{2}}{cr} \hat{\Lambda}^{2} \psi_{n_{r},j, m_{j}}^{(u)} + \left( -\hat{p}_{r} \hat{\Lambda}^{2} + \frac{i}{r} \hat{\Lambda}^{3} + imZe^{2} \hat{\Lambda} \right) \hat{\sigma}_{r} \psi_{n_{r},j, m_{j}}^{(d)} = \epsilon \psi_{n_{r},j, m_{j}}^{(d)} .
\end{array}
\] 
Substituting (\ref{anzatsp}) and using explicit expressions for radial functions (\ref{F^+,-;G^+,-}),  we see that the bispinor (\ref{Bispin_gen}) is the eigen bispinor of the BELI provided free constants $ A $ and $ B $ satisfy the relations 
\begin{equation}
\label{EqC_1,2_BEL} 
\begin{array}{c}
\left( n_{r} + 2\gamma_{j} \right) A + \frac{\epsilon \mathcal{N}_{n_{r},j}}{Z me^{2} \hbar \kappa_{j} } B = 0 ,\\
\frac{\epsilon \mathcal{N}_{n_{r},j}}{Z me^{2} \hbar \kappa_{j} } A + n_{r} B = 0 .
\end{array}
\end{equation}
Respectively, the condition for non-trivial values of these constants determines two eigenvalues $ \epsilon_{BEL} = \pm \hbar \kappa_{j} a_{n_{r},j} $, where $ \kappa_{j} $ and $ a_{n_{r},j} $ are defined by Eqs. (\ref{kappa_j}) and (\ref{eig_JL}), respectively. Then from Eq. (\ref{EqC_1,2_BEL}), one finds two solutions for constants which lead to two orthonormal bispinors, corresponding to two independent solutions for numbers $ (\beta_{1},\beta_2) $ (cp. (\ref{beta_D}) and (\ref{beta_JL}))
\begin{equation}
\label{beta_BEL} 
\left(\begin{array}{c}
\beta_{1}^{(+)} \\
\beta_{2}^{(+)}
\end{array} \right) = \left(\begin{array}{c}
(1/\sqrt{2}) \\
- (1/\sqrt{2}) 
\end{array} \right)  , \quad \left(\begin{array}{c}
\beta_{1}^{(-)} \\
\beta_{2}^{(-)}
\end{array} \right) = \left(\begin{array}{c}
(1/\sqrt{2})  \\
(1/\sqrt{2}) 
\end{array} \right) ,
\end{equation}
one of which, $ \beta_{\nu}^{(+)} $, corresponds to the states with positive eigenvalue of BELI, $ \epsilon_{BEL} = \hbar \kappa_{j} a_{n_{r},j} $, and the other one, $ \beta_{\nu}^{(-)} $, -- with negative, $ \epsilon_{BEL} = - \hbar \kappa_{j} a_{n_{r},j} $. The eigenstates, as before, are characterized by four quantum numbers $ (n_{r}, j, m_{j},\sigma=\pm ) $.

\section{General solution of the Dirac equation}

Each set $ \lbrace \hat{H}, \hat{J^{2}}, \hat{J}_{z}, \hat{{\cal I}}_{inv} \rbrace $, where $ \hat{{\cal I}}_{inv} $ is $ \hat{K}_D $, $ \hat{A}_{JL} $ or $ \hat{{\cal I}}_{BEL} $, results in its own system of eigen bispinors with certain values of the parameters $ \beta_{\nu} $ in the general solution (\ref{Bispin_gen}). On the other hand, such a solution can be sought as a single eigen bispinor of the set with generalized invariant 
\begin{equation}
\label{Inv_g} 
\hat{{\cal I}}_{inv} = \hat{{\cal I}}_{gen} = \frac{c_{D}}{\hbar} \hat{{\cal I}}_{D} + \frac{c_{JL}}{mZe^{2}}  \hat{{\cal I}}_{JL} + \frac{c_{BEL}}{\hbar mZe^{2}} \hat{{\cal I}}_{BEL} ,
\end{equation}
which is given by arbitrary linear combination  with coefficients $ c_{D} $, $ c_{JL} $ and $ c_{BEL} $ determining the  weight of each invariant contribution. 

Equality $ \hat{{\cal I}}_{gen} \Psi_{n_{r}, j, m_{j}} = \epsilon_{gen} \Psi_{n_{r}, j, m_{j} } $ leads to the following equations for constants $ A $ and $ B $:
\begin{equation}
\label{A-B_gen} 
\begin{array}{c}
\left( n_{r} + 2\gamma_{j} \right) A + \frac{\left( \epsilon + \kappa_{j} c_{D} \right) \mathcal{N}_{n_{r},j} }{\kappa_{j}c_{BEL} - ic_{JL}} B = 0 , \\
\frac{\left( \epsilon - \kappa_{j} c_{D} \right) \mathcal{N}_{n_{r},j} }{\kappa_{j}c_{BEL} + ic_{JL}} A + n_{r} B = 0 .
\end{array}
\end{equation}
The condition for their non-trivial solution defines the eigenvalues 
\begin{equation}
\label{eps_gen} 
\epsilon_{gen} = \pm \epsilon_{n_{r},j} , \quad \epsilon_{n_{r},j} = \sqrt{\kappa_{j}^{2} c_{D}^{2} + \left( c_{JL}^{2} + \kappa_{j}^{2} c_{BEL}^{2} \right) \frac{n_{r} \left( n_{r} + 2\gamma_{j} \right)}{\mathcal{N}_{n_{r},j}^{2}}} , 
\end{equation} 
and Eqs.(\ref{A-B_gen}) admit two solutions which correspond to positive and negative eigenvalues of the generalised invariant (\ref{Inv_g}). Introducing angles $ \theta $ and $\phi $ via relations
\begin{equation}
\label{theta-phi}
\cos \theta = \frac{\kappa_{j} c_{D}}{\epsilon_{n_{r},j}},\quad  \tan \phi = \frac{c_{JL}}{\kappa_{j}c_{BEL}} , \quad  \kappa_{j}c_{BEL} + ic_{JL} = \sqrt{c_{JL}^{2} + \kappa_{j}^{2}c_{BEL}^{2} } e^{i\phi} ,
\end{equation}
one comes to two solutions for the generalized parameters $ \beta_{\nu} $ 
\begin{equation}
\label{beta_nu,sigma} 
\left(\begin{array}{c}
\beta_{1}^{(+)} \\
\beta_{2}^{(+)}
\end{array} \right) = \left(\begin{array}{c}
e^{i\phi /2} \cos \frac{\theta}{2}  \\
-e^{-i\phi /2} \sin \frac{\theta}{2} 
\end{array} \right)  , \quad \left(\begin{array}{c}
\beta_{1}^{(-)} \\
\beta_{2}^{(-)}
\end{array} \right) = \left(\begin{array}{c}
 e^{i\phi /2} \sin \frac{\theta}{2}\\
 e^{-i\phi /2} \cos \frac{\theta}{2}
\end{array} \right) ,
\end{equation} 
where the first pair $ \beta_{\nu}^{(+)} $ corresponds to the states with positive eigenvalue of the invariant (\ref{Inv_g}), $ \epsilon_{gen} = \epsilon_{n_{r},j} $, and the second one, $ \beta_{\nu}^{(-)} $, -- with negative, $ \epsilon_{gen} = - \epsilon_{n_{r},j} $. 

Therefore, solution of the last eigenvalue equation in (\ref{state}) defines two orthogonal bispinors (\ref{Bispin_gen}), in which parameters $ \beta_{\nu}^{(\pm)} $ are determined by  expressions (\ref{beta_nu,sigma}) with angles $ \theta $ and $ \phi $ given in (\ref{theta-phi}). In particular cases when only one of the three constants $ c_{D} $, $ c_{JL} $ or $ c_{BEL} $ is non-zero, coefficients $ \beta_{\nu}^{(\pm)} $ are reduced to (\ref{beta_D}), (\ref{beta_JL}) or (\ref{beta_BEL}), respectively. In all cases the eigen bispinors $ \Psi_{n_{r}, j, m_ {j}, \sigma} (\mathbf {r}) $ for the DE are characterized by the set of quantum numbers where $ \sigma $ takes two values in accordance with the sign of the eigenvalue of the corresponding spin invariant operator. 

Note that the structure of bispinors (\ref{bispin^D}) is relatively simple as compared with the rest ones, and the general solution (\ref{Bispin_gen}) can be represented in the form of the linear combination of the Darwin solutions 
\begin{equation}
\label{eigbispi_g} 
\Psi_{n_{r}, j, m_{j},\sigma } (\mathbf{r})= \beta_1^{(\sigma)} \Psi_{n_{r}, j, m_{j},+ }^{(D)}(\mathbf{r}) + \beta_2^{(\sigma)} \Psi_{n_{r}, j, m_{j},- }^{(D)} (\mathbf{r}) .
\end{equation}

All sets for four independent operators include the invariant $\hat{J}^2$ and differ by spin invariants only. Therefore, the DE can be exactly solved by variable separation in spherical coordinates.  Obviously, the eigen bispinor of any of the invariants can be expanded over the eigen bispinors of another one. Due to anticommutation relations between spin invariants, an interconverting transformation includes only two bispinors corresponding to different signs of $ \sigma $ \cite{Dahl}, and expression (\ref{eigbispi_g}) is expansion of eigen bispinors of $ \hat{{\cal I}}_{gen} $ over the eigen bispinors of the DI.  

\section{Classification of quantum states }

It is worth to note that in spite of several solutions, the eigen spectrum of the Dirac Hamiltonian with Coulomb interaction does not depend on the choice of the set $ \lbrace \hat{H}_{D}, \hat{J}_{z}, \hat{J^{2}}, \hat{{\cal I}}_{inv} \rbrace $. Indeed, electron states in this potential are defined by the full set of quantum numbers $ \lbrace n_{r}, j, m_{j}, \sigma \rbrace $, where $ \sigma = \pm $. Instead of this, it is more convenient to use the principal quantum number \cite{BetSol} $ n= n_{r} + \kappa_{j} = 1,2,... $ and to characterize stationary states by the set of quantum numbers $ \lbrace n, j, m_{j}, \sigma \rbrace $ where the principal quantum number $ n $ takes  values $ n = 1,2,\ldots $, quantum number of the total angular momentum $ j $ takes semi-integer values in the range $ 1/2 \leq j \leq n - 1/2 $ ($ 1 \leq \kappa_{j} \leq n $), magnetic quantum number $ m_{j} $ takes positive and negative semi-integer values in the range $ -j \leq m_{j} \leq j $ and $ \sigma $, in general, takes two values. 

In this case the energy levels of the hydrogen spectrum, calculated from Eq. (\ref{E_rel2}) with the accuracy of the lowest order with respect to the fine structure constant, take the form 
\begin{equation}
\label{E_n,j} 
E_{n,j} - mc^{2} = W_{n,j} \simeq -W_{R} \frac{Z^{2}}{n^{2}} \left( 1 + \frac{4n - 3\kappa_{j}}{4n^{2}\kappa_{j}} Z^{2} \alpha^{2} \right) ,
\end{equation}
where $ W_{R} = e^{4}m/2\hbar^{2} $ is the \textit{Rydberg energy} and number $ \kappa_{j}  $ was defined in Eq. (\ref{kappa_j}).  

Sub-levels with the maximal possible for the given principal quantum number, $ j = j_{max} = n - 1/2 $, are very special in the fine structure. At these levels when $ \kappa_{j_{max}} = n $, the polynomials of the order $ n_{r} = n - \kappa_{j_{max}} = 0 $ become constant and according to definitions (\ref{tildP,Q}), polynomials $ P_{0,j}^{(+)} $ and $ Q_{0,j}^{(-)} $  are non-zero, while another two polynpmials vanish, $ P_{0,j}^{(-)} = Q_{0,j}^{(+)} \equiv  0 $. In these states the eigenvalues of $ \hat{A}_{JL} $ and $ \hat{{\cal I}}_{BEL} $ are also zero, the sign of $ \sigma = \pm $ is meaningless, and the states on this level are described by the only bispinor $ \Psi_{n, j=n-1/2, m_{j},+ }^{(D)} $ in (\ref{bispin^D}), while $ \Psi_{n, j=n-1/2, m_{j},- }^{(D)} = 0 $. 

Thus, the states with $ n_{r} = 0 $ are described by bispinors 
\begin{equation}
\label{Bisp_n=0} 
\Psi_{n, j_{max}, m_{j},+ } (\mathbf{r}) = \left( \frac{2Z}{r_{B}\mathcal{N}_{0,j}} \right)^{3/2} C_{0,j} e^{-\frac{r_{0,j}}{2 }} r_{0,j}^{\gamma_{j}-1}
\left( \begin{array}{c} 
 \chi_{j-1/2, M,+}\left(\vartheta ,\varphi \right)  \\ 
-\sqrt{\frac{1 - \varepsilon_{0,j}}{1 + \varepsilon_{0,j}}} \chi_{j+1/2, M,-} \left(\vartheta ,\varphi \right)    
\end{array} \right) ,
\end{equation}
which do not depend on the choice of the eigen bispinors system and $ \sigma $ takes one value only. The corresponding energy levels of the fine structure are $ 2n $ times degenerate in accordance with the magnetic quantum number $ m_{j} = \pm 1/2, \ldots ,\ \pm (n-1)/2 $. 

The rest $ n-1 $ sub-levels (at $ n \geq 2 $, $ j < (2n-1)/2 $) of the fine structure of each multiplet are described by spinors (\ref{Bispin_gen}) with polynomials of the order $ n_{r} = n - \kappa_{j} \geq 1 $, in which $ m_{j} $ takes $ 2(n-n_{r}) $ values, and $ \sigma $ can have both signs. The sub-level of the fine structure is degenerate with respect to the magnetic quantum number $ m_{j} $ and spin number $ \sigma $ and, therefore, is $ 4(n-n_{r}) $ times degenerate. This means that $ 2n + 4\sum_{n_{r}=1}^{n-1} \left( n-n_{r} \right) = 2n^{2} $ states correspond to the same principal quantum number $ n $, as it should be. 

Using the found   bispinors $ \Psi_{n, j, m_{j},\sigma } (\mathbf{r}) $, it is possible to calculate the distribution of probability amplitude for the atomic orbitals 
\[
w_{n, j, m_{j},\sigma} = \Psi_{n, j, m_{j},\sigma }^{\dagger} (\mathbf{r}) \Psi_{n, j, m_{j},\sigma } (\mathbf{r}) = \mid \psi^{(u)}_{n, j, m_{j},\sigma} \mid^{2} + \frac{1 - \varepsilon_{n,j}}{1 + \varepsilon_{n,j}} \mid \psi^{(d)}_{n, j, m_{j},\sigma} \mid^{2} ,
\]
and their electron spin orientation  
\[
\Psi_{n, j, m_{j},\sigma }^{\dagger} (\mathbf{r}) \bm{\hat{\Sigma}} \Psi_{n, j, m_{j},\sigma } (\mathbf{r}) = \left. \psi^{(u)}\right.^{\dagger}_{n, j, m_{j},\sigma} \bm{\hat{\sigma}} \psi^{(u)}_{n, j, m_{j},\sigma} + \frac{1 - \varepsilon_{n,j}}{1 + \varepsilon_{n,j}} \left. \psi^{(d)}\right.^{\dagger}_{n, j, m_{j},\sigma} \bm{\hat{\sigma}} \psi^{(d)}_{n, j, m_{j},\sigma}. 
\]

In view of the smallness of the parameter $ Z\alpha \ll 1 $, contribution of the lower spinor of the order $ |Z\alpha |^2 \ll 1 $ in these expressions is negligibly small and  can be ignored. By the same reason the upper spinor can be expanded with respect to the small parameter, and terms $ \sim Z^{2}\alpha^{2} $ can be omitted also. Therefore, in calculations of the mean-values the lower spinor can be omitted. The upper spinor has the form 
\begin{equation}
\label{psi_u} 
\psi_{n,j,m_{j},\sigma}^{(u)} \simeq  \beta_1^{(\sigma )} R_{n,j}^{(+)}\left( r \right) \chi_{j-1/2, m_{j},+}\left(\vartheta ,\varphi \right) + \beta_2^{(\sigma )}  R_{n,j}^{(-)}\left( r \right) \chi_{j+1/2, m_{j},-} \left(\vartheta ,\varphi \right)  ,
\end{equation}
where numbers $ \beta_{1,2}^{(\sigma)} $ are given in (\ref{beta_nu,sigma}) and $ R_{n,j}^{(\pm)}\left( r \right) $ are the radial functions 
\[
R_{n,j}^{(\pm)}\left( r \right) = \left(\frac{2Z}{n r_{B}} \right)^{3/2} C_{n,j} e^{-\rho_{n}/2} \rho_{n}^{\gamma_{j} - 1} P_{n,j}^{(\pm)}(\rho_{n}),
\]
with the dimensionless radial variable $ \rho_{n} = 2Zr/nr_{B} $ (see (\ref{x_n,j})). Here the normalization constant  and the polynomials in the limit  $ Z\alpha \rightarrow 0 $ are defined in Eq. (\ref{cnrj}) and (\ref{tildP,Q}), respectively. In this approximation the function $ R_{n,j}^{(\pm)} $ coincides with the Schr\"{o}dinger radial function except one difference: at $ j=1/2 $ the value $ \gamma_{j} $ is slightly less than $ 1 $ which leads to a small negative power of $ r $. Of course, all integrals connected with the spectrum, are convergent. In fact, this corresponds to the non-relativistic approximation when an electron is described by the Pauli spinor (\ref{psi_u}). But it is necessary to keep in mind that excitations of the Dirac spinor field (particles and antiparticles) are described by the full 4-component field amplitude (bispinor) $ \Psi_{n, j, m_{j},\sigma } $. 

So, to calculate the density distribution and the spin direction in the corresponding eigenstate, one can use (\ref{psi_u}) as the Pauli spinors, $ \psi_{n,j,m_{j},\sigma} = \psi_{n,j,m_{j},\sigma}^{(u)} $, which gives
\begin{equation}
\label{w,s_nr} 
w_{n, j, m_{j},\sigma} \simeq \mid \psi_{n, j, m_{j},\sigma} \mid^{2 } , \quad \mathbf{s}_{n, j, m_{j},\sigma} \simeq \frac{\psi^{\dagger}_{n, j, m_{j},\sigma} \bm{\hat{\sigma}} \psi_{n, j, m_{j},\sigma}}{w_{n, j,m_{j},\sigma}} .
\end{equation}

Electron states are characterised by the full set of quantum numbers $ \lbrace n,j,m_{j},\sigma \rbrace $. In states (\ref{bispin^D}) with the definite value of the DI with constants $ \beta_{1,2}^{(\sigma)} $ being fixed in Eq. (\ref{beta_D})), the upper spinor of bispinors $ \psi_{n,j,m_{j},\sigma}^{(u)} = R_{n,j}^{(\pm)} \chi_{l, m_{j},\pm} $ involves spinors with integers $ l = 0,1, 2, \ldots $. In view of the smallness of the lower spinor $ \sim |Z\alpha |^2 $, the conventional spectroscopic classification of  states by $ s , \, p,\,d,\, \ldots $ based on number $ l $, is used. But in our theory $ l $ defines the order of spherical harmonics in the main spinor and is not connected with the angular momentum operator $ \hat{L}^{2} $ because angular momentum is no longer an integral of motion.

The states on the levels with the maximal value of $ j $ at the given principal quantum number $ n $, i.e., at $ \kappa_{j} = n $, are special and differ from the rest ones. The ground state $ 1S_{1/2} $ and states on the upper sub-levels in the fine structure multiplets $ 2P_{3/2}, 3D_{5/2},\ldots $ belong to such states. They are described by bispinors (\ref{Bisp_n=0}) and, correspondingly, by the Pauli spinor (\ref{psi_u}) in the non-relativistic approximation, in which $ \beta_{1} = 1 $ and $ \beta_{2} = 0 $. In these states the probability density andspin orientation in the electron cloud do not depend on the choice of the spin invariant. For example, for the two-fold degenerate ground states $ 1S_{1/2} $ ($ n=1,j=1/2,m_{j}=\pm 1/2, \sigma =+ $) one has $ \psi_{1,1/2,\pm 1/2,+} = \psi_{1S_{1/2}} $,
\begin{equation}
\label{Psp_1} 
\begin{array}{c}
\psi_{1S_{1/2}} \left( \mathbf{r} \right) = R_{10} \chi_{0, \pm 1/2,+} , \quad R_{10} \simeq \left( Z/r_{B} \right)^{3/2} 2 e^{-Zr/r_{B}} \left( 2Zr/r_{B} \right)^{\gamma_{1/2}-1} , \\
\chi_{0, \frac{1}{2},+} = \frac{1}{\sqrt{4\pi}} \left( \begin{array}{c} 1 \\
0
\end{array} \right) , \quad \chi_{0, -\frac{1}{2},+} = \frac{1}{\sqrt{4\pi}} \left( \begin{array}{c} 0 \\
1
\end{array} \right) .
\end{array}
\end{equation}
For the four-fold degenerate level $ 2P_{3/2} $ ($ n=2,j=3/2,m_{j}=\pm 1/2,\pm 3/2, \sigma =+ $) the states are described by spinors $ \psi_{2,3/2,m_{j},+} = \psi_{2P_{3/2}} $,
\begin{equation}
\label{Psp_2,3/2} 
\begin{array}{c}
\psi_{2P_{3/2}} \left( \mathbf{r} \right) = R_{20} \chi_{1, m_{j},+} , \quad R_{20} \simeq \left( Z/r_{B} \right)^{3/2} (1/2\sqrt{6}) e^{-Zr/2r_{B}} \frac{Zr}{r_{B}}  , \quad m_{j} = \pm \frac{1}{2} , \, \pm \frac{3}{2} , \\ 
\chi_{1, \frac{1}{2},+} = \frac{i}{\sqrt{8\pi}} \left( \begin{array}{c} 2 \cos \vartheta \\
-e^{i\varphi} \sin \vartheta
\end{array} \right) , \quad \chi_{1, -\frac{1}{2},+} = \frac{i}{\sqrt{8\pi}} \left( \begin{array}{c} e^{-i\varphi} \sin \vartheta \\
2 \cos \vartheta
\end{array} \right) , \\
\chi_{1, \frac{3}{2},+} = -i \sqrt{\frac{3}{8\pi}} e^{i\varphi} \sin \vartheta \left( \begin{array}{c} 1 \\
0
\end{array} \right) , \quad \chi_{1, -\frac{3}{2},+} = i \sqrt{\frac{3}{8\pi}} e^{-i\varphi} \sin \vartheta \left( \begin{array}{c} 0 \\
1
\end{array} \right) . 
\end{array}
\end{equation}

It follows that the angular spatial dependence of the density distribution and  spin orientation of an electron on these levels are described by  spherical spinors $ \chi_{l, m_{j},+} $. In the states with maximal value of magnetic quantum number $ m_{j} = \pm j $ the spin is directed along or opposite to the polar axis depending on the sign of $ m_{j} $. At $ \vert m_{j} \vert < j $ the spin orientation in electron cloud depends on polar angle $ \vartheta $ being oppositely directed for opposite signs of $ m_{j} $. For instance, in the ground state $ 1S_{1/2} $ the probability density distribution is spherically symmetric with two possible spin orientation $  \mathbf{s}_{1,1/2,\pm 1/2} = \pm \mathbf{e}_{z} $. In the $ 2P_{3/2} $ state we have 
\begin{equation} 
\label{2,3/2}
\begin{array}{c}
w_{2,3/2} (\mathbf{r}) = \frac{3}{8\pi} R_{20}^{2}\left(r \right) \sin^{2} \vartheta ; \quad \mathbf{s}_{2,3/2,\pm 3/2} = \pm \mathbf{e}_{z}, \\
w_{2,3/2} (\mathbf{r}) = \frac{1}{8\pi} R_{20}^{2}\left(r \right) \left( 1 + 3 \cos^{2} \vartheta \right) , \; \mathbf{s}_{2,3/2,\pm 1/2} = \pm \frac{ \mathbf{e}_{z} \left( 5\cos^{2} \vartheta - 1 \right) - \mathbf{e}_{\rho} 2\sin 2\vartheta }{1 + 3\cos^{2} \vartheta} .
\end{array}
\end{equation}
Worth mentioning is a peculiarity of the states on these levels which complete population of the $ n $-th shell of the hydrogen-like spectrum. Hartree D.R. \cite{Hartree} has shown that summing of probabilities $ w_{n,j= n-1/2, m_{j}} $ by $ m_{j} $ from $ m_{j} = +1/2 $ up to $ m_{j} = +j $ gives spherically symmetric charge distribution. In particular, one can see from Eq. (\ref{2,3/2}) that $ w_{2,3/2,\pm 1/2} + w_{2,3/2,\pm 3/2} = R_{20}^{2}\left(r \right)/2\pi $. 

On the other sub-levels of the multiplet fine structure  electron states are described by quantum numbers $ m_{j} $ and $ \sigma $ which take two values. In these states the probability density and spin orientation significantly depend on the choice of the spin invariant in the set $ \lbrace \hat{H}, \hat{J^{2}}, \hat{J}_{z}, \hat{{\cal I}}_{inv} \rbrace $. For the states with definite value of the DI, the four-fold degenerate level $ E_{2,1/2} $ includes $ 2S_{1/2} $ states ($ n=2,j=1/2,m_{j}=\pm 1/2, \sigma = + $) which are described by spinors
\begin{equation}
\label{st2S_1/2} 
\psi_{2S_{1/2}}\left( \mathbf{r} \right) = R_{20} \chi_{0,\pm 1/2,+} , \quad R_{20} \simeq \left(\frac{Z}{r_{B}} \right)^{3/2} e^{-Zr/2r_{B}} \frac{1}{\sqrt{2}} \left( 1 - \frac{Zr}{2r_{B}} \right) \left( \frac{Zr}{r_{B}} \right)^{\gamma_{1/2}-1} , 
\end{equation}
and $ 2P_{1/2} $ states ($ n=2,j=1/2,m_{j}=\pm 1/2, \sigma = - $) for which 
\begin{equation}
\label{st2P_1/2} 
\begin{array}{c}
\psi_{2P_{1/2}}\left( \mathbf{r} \right) = R_{21} \chi_{1,\pm 1/2,-} , \quad R_{21} \simeq - \left(\frac{Z}{r_{B}} \right)^{3/2} \frac{1}{2\sqrt{6}} e^{-Zr/2r_{B}} \frac{Zr}{r_{B}} \left( \frac{Zr}{r_{B}} \right)^{\gamma_{1/2}-1} ,  \\
\chi_{1, \frac{1}{2},-} \left( \vartheta ,\varphi \right) = \frac{-i}{\sqrt{4\pi}} \left( \begin{array}{c} \cos \vartheta \\
e^{i\varphi} \sin \vartheta
\end{array} \right) , \quad \chi_{1, -\frac{1}{2},-} \left( \vartheta ,\varphi \right) = \frac{i}{\sqrt{4\pi}} \left( \begin{array}{c} -e^{-i\varphi} \sin \vartheta \\
\cos \vartheta
\end{array} \right) . 
\end{array}
\end{equation}
In Ref. \cite{White} the probability amplitude for Darwin solution is given in graphical presentation and it is shown that radial distributions, described by Shr\"odinger and Dirac functions, are almost identical. This means that the approximate expression via the Pauli spinors (\ref{psi_u}) reproduces $ \vert \Psi_{n, j, m_{j},\sigma } \vert^{2} $ with high accuracy. 

For sub-level $ E_{2,1/2} $ we have 
\[
\begin{array}{c}
w_{2S_{1/2}} = \frac{1}{8\pi} \vert R_{20}\left(r \right) \vert^{2}  , \quad \mathbf{s}_{2S_{1/2},\pm 1/2 } = \pm \mathbf{e}_{z} , \\
w_{2P_{1/2}} = \frac{1}{8\pi} \vert R_{21}\left(r \right) \vert^{2}  , \quad \mathbf{s}_{2P_{1/2},\pm 1/2 } = \pm \left(\mathbf{e}_{r} \cos \vartheta  +\mathbf{e}_{\vartheta}  \sin \vartheta \right)
\end{array}
\]
where $ \mathbf{e}_{r} $ and $ \mathbf{e}_{\vartheta} $ are unit vectors in spherical coordinate system. It is seen, as it had been pointed out by H. E. White \cite{White}, "not only are all $ S $ states spherically symmetrical, as on the Schr\"odinger theory, but also one electron in a $ 2P_{1/2} $~ state". 

Nonetheless, the spin invariant itself is not determined \textit{a priori} and an electron can, at least from the point of view of its energy, be in any of the discussed above states. Therefore, electron states should be described by the general solution (\ref{Bispin_gen}) or (\ref{eigbispi_g}) (or by Pauli spinors (\ref{psi_u})) in the non-relativistic limit). 

This means that on sub-levels with $ j<n-1/2 $ of the fine structure multiplet, the shape of electron cloud and the spin orientation in it take different forms. The deformation of electron cloud and changes in spin orientation are mutually connected and are controlled by variation of the values of $ \beta_{1,2}^{(\sigma)} $, i.e., by parameters $ \theta $ and $ \phi $ in (\ref{beta_nu,sigma}, that characterize  spin degree of freedom in the spinor (\ref{psi_u}). In all cases the states with positive and negative values of the magnetic number $ m_{j} $ have the same probability density distribution, but opposite spin orientations. For example, the states on sub-level $ E_{2,1/2} $ are defined by Pauli spinors 
\begin{equation}
\label{psi_2,1/2} 
\psi_{2,1/2,m_{j},\sigma}^{(g)} =  \beta_1^{(\sigma )} R_{2,1/2}^{(+)}\left( r \right) \chi_{0, m_{j},+} + \beta_2^{(\sigma )} R_{2,1/2}^{(-)}\left( r \right) \chi_{1, m_{j},-} \left(\vartheta ,\varphi \right) ,
\end{equation} 
where the values of coefficients $ \beta_{1,2}^{(\sigma )} $  for two states with different values of  $ \sigma $ are defined in Eq. (\ref{beta_nu,sigma}). With account of the relations (\ref{st2S_1/2})-(\ref{st2P_1/2}) the spinor (\ref{psi_2,1/2}) can be re-written in the form $\psi_{2,1/2,m_{j},\sigma} =  \beta_1^{(\sigma )} \psi_{2S_{1/2}} + \beta_2^{(\sigma )} \psi_{2P_{1/2}}  $ which reflects $ sp $-hybrydization of the states on this level.

Figures \ref{fig:1}--\ref{fig:6}  below show the probability and charge density distributions for the states $ E_{2S_{1/2}-2P_{1/2}} $. They demonstrate how the probability changes depending on the spin invariant. In the case of JLI the symmetry with respect to reflection from the $ xy $-plane perpendicular to the polar $z-$axis and, respectively, spherical symmetry are broken, charge density shifts along the polar axis in one or opposite direction depending on $\sigma $ (see Fig. \ref{fig:5}). Spin orientations $ \mathbf{s}_{2, 1/2, m_{j},\sigma} $ (\ref{w,s_nr}) for each invariant are given by different expressions from which it follows that spin spatial dependencies differ significantly, as it is also clearly seen from the figures. Moreover, it follows from the above obtained expressions that for the states $n=2, j=1/2, m=\pm 1/2$ corresponding to the DI and JLI, spins lie in the plane $(z, x)$, while in the case of BELI  spins lie in this plane only at $\vartheta =0, \pi $, at other angles spin orientations have a component in $\pm \mathbf{e}_{\varphi} $ direction. This is shown in Fig.  \ref{fig:6} for the state with $m=1/2,\, \sigma = +$ by arrows with symbols ${ \bullet}$ and $\mathbf{\times}$, respectively.  
In the state $m=-1/2,\, \sigma = +$ spins satisfy the relation $\mathbf{s}^{BEL}_{2, 1/2, -1/2,+} = - \mathbf{s}^{BEL}_{2, 1/2, 1/2,+} $, and, thus, they have opposite orientation to the one shown in Fig. \ref{fig:6} (c). In the state $m=1/2,\, \sigma = -$ spins have orientation in the plane $(z,x)$ like the one, shown in Fig. \ref{fig:6} (c), but have opposite $\mathbf{e}_{\varphi} $ component out of the plane $(z,x)$. In the other state spin orientation satisfies the relation $\mathbf{s}^{BEL}_{2, 1/2, -1/2,-} = - \mathbf{s}^{BEL}_{2, 1/2, 1/2,-} $.

\section{Final remarks} 

From the above studies it follows that electron stationary states at the levels of hydrogen-like energy spectrum are determined by the full set of quantum numbers\footnote{In fact, it follows from the DE that there is another quantum number which indicates the sign of the Dirac Hamiltonian eigenvalue (particle and antiparticle). The present technologies already allow studying of the fine structure of antihydrogen \cite{ALPHA_Col}.} $ \lbrace n, j, m_{j}, \sigma \rbrace $. It has been shown that the eigenvalue equation $ \hat{{\cal I}}_{gen} \Psi_{n_{r}, j, m_{j}} = \epsilon_{gen} \Psi_{n_{r}, j, m_{j} } $ fixes free parameters in the general solution (\ref{eigbispi_g}) and defines the two orthogonal eigen bispinors corresponding to the two signs $ \sigma = \pm $ of the eigenvalue $ \epsilon_{gen} $. As long as the free parameters in the general solution of the DE are not fixed, the spin vector is not defined. The free parameters and spin vector  are determined by the choice of the particular invariant  $ \hat{K}_D $, $ \hat{A}_{JL} $ or $ \hat{{\cal I}}_{BEL} $. As a result, the spinors in the eigen bispinor become unambiguously determined and in this sense these invariants can be called "spin invariants" and $ \sigma $ -- "spin quantum number" according to conventional meaning of this number.

Three quantum numbers $ n, j, m_{j} $ correspond to spatial degree of freedom and $ \sigma $ to the spin one. In the non-relativistic limit electron is described by two-component Pauli spinor and $ \sigma $ is usually connected with the spin direction. Such statement is valid for systems with the translational symmetry when the  integrals of motion are given by components of the momentum (quasi-momentum). In the central field when a constant of motion is $ \hat{J}_{z} $, the spin direction is determined by the sign of the \textit{magnetic quantum number}.   

Due to the hidden symmetry the energies (\ref{E_rel2}) are degenerate with respect to magnetic $ m_{j} $ and spin $ \sigma $ quantum numbers. External perturbations can  break some symmetries and remove the degeneracy. For instance, an external magnetic field removes the degeneracy with respect to $ m_{j} $.  Zeeman effect causes splitting of the corresponding states, in particular the $ 2P_{3/2} $ state splits into four sub-levels, whereas each state $ 2S_{1/2} $ and $ 2P_{1/2} $ splits into two sub-levels. Some external factors (e.g., the interaction with quantum fluctuations of the vacuum electromagnetic field) remove the degeneracy with respect to $ \sigma $ and cause  hyperfine splitting (the Lamb shift) of states with different $ \sigma $: splitting between the states $ 2S_{1/2} $ ($ \sigma = - $) and $ 2P_{1/2} $ ($ \sigma = + $). The states of fine-structure multiplets with the maximal value $ j $ at the given principal quantum number are characterized by one value of the quantum number $ \sigma $ only  and, therefore, stay unsplitted. 

Finally, note, the expectation values can be calculated within the non-relativistic approximation with replacement of Dirac bispinors by Pauli spinors (\ref{psi_u}). Such calculations give the results which with high accuracy coincide with the exact results, and nevertheless, the important thing is that even in the non-relativistic approximation the electron is described by Dirac bispinor.  Even though the spinors of this bispinor have different weights, neither of them can be ignored, and the complete description of an electron states requires both spinors. The DI has a diagonal block matrix structure, and the non-relativistic corrections are present in the lower spinor only. Its account results in the appearance in the non-relativistic approximation of the relativistic correction corresponding to the Thomas term,  which in the case of the central field is known as spin-orbit interaction proportional to the product $  \bm{\sigma} \cdot \bm{L}$.  If the matrix of a spin invariant is not diagonal, as it takes place for spin invariants  $\hat{{\cal I}}_{JL}$ and  $\hat{{\cal I}}_{BEL}$, relativistic corrections are present in both spinors, which results in  another form of the spin-orbit interaction in the non-relativistic approximation. For example, as it can be easily shown,  for the JLI the spin-orbit interaction operator acquires different form, namely, it is proportional to the product $  \bm{\sigma} \cdot \bm{A}_+ $ (see (\ref{opPi})).  Another example of non-conventional form of this interaction is quasi-two-dimensional electron gas \cite{LTP}.  

Present  study demonstrates spin state variability in the hydrogen-like spectrum. Since energies of the different spin states are degenerate, a small external factor, e.g., the one breaking the inversion symmetry, can be enough to cause electron transition from one spin orientation to another one. In this respect we recall that new technologies such as spintronics, spin chemistry etc., have attained a large amount of new data regarding the role of spin and its manifestation in various processes. Spin based effects take place in  spincaloritronics devices,  heat-to-spin conversion and energy harvesting systems, etc.,  spin-orbit torque devices for electric magnetization switching, as well as in the  design of quantum materials \cite{Naaman,Michaeli1,Chernyshov}, spin relativistic chemistry \cite{Rosenberg,Bustami}. Precision measurements of  spin related 
fine structure and Lamb shift in the anti-hydrogen spectrum have been performed \cite{ALPHA_Col}. They  can be used  as tests of the charge–parity–time symmetry, as method of the determination of other fundamental quantities of such systems, etc. 

Another field of spin sensitive processes is biological systems, in which electrons play essential role not only in the  storage and transport of energy and information, but also in biorecognition connected with  spin selectivity and  chiral-induced spin selectivity. For instance, electron transfer through diamagnetic material (macromolecules, DNA) results in  specific spin states of electrons  \cite{EL,Michaeli,Varade,Zollner} which is attributed to spin-orbit interaction. This underlines importance of further investigation of spin effects in different systems.

\begin{figure}[t]
\vspace{4cm}
\begin{picture}(16,8)
  \includegraphics[width=4cm]{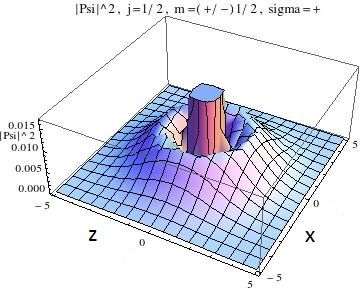}
   \includegraphics[width=3.5cm]{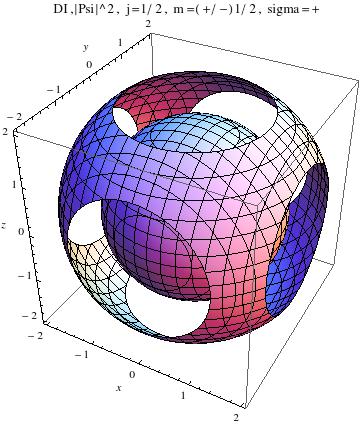}
   \includegraphics[width=4cm]{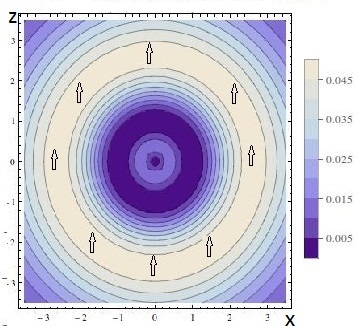}
  \end{picture}
  \caption{(a) -- Spatial  distribution of electron probability  described by the DE with the DI for the  states $n=2,\, j=1/2,\, m=\pm 1/2,\, \sigma =+$  in the plane $(x,z)$. (b) -- A pattern of two 3D distributions of  probabilities equal to the value 0.005, in the Cartesian coordinate system.  (c)  -- Charge 'isodensity' lines in the plane $(x,z)$, arrows show spin orientation in the area of the highest charge density for the state $m=1/2$. For the state $m=-1/2,\, \sigma =+$ spins have opposite orientation. 
}
  \label{fig:1}
\end{figure}

\begin{figure}[t]
\vspace{4cm}
\begin{picture}(16,8)
  \includegraphics[width=4cm]{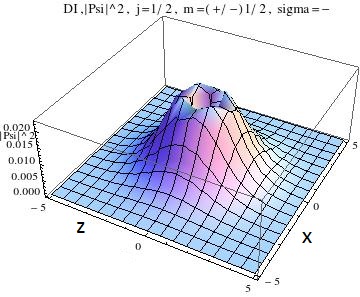}
   \includegraphics[width=3.5cm]{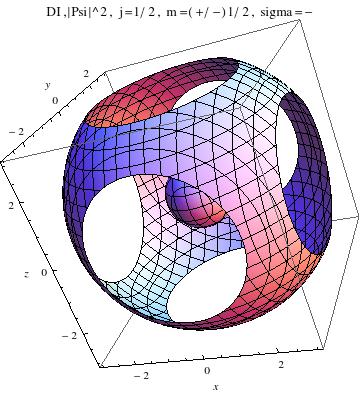}
    \includegraphics[width=4cm]{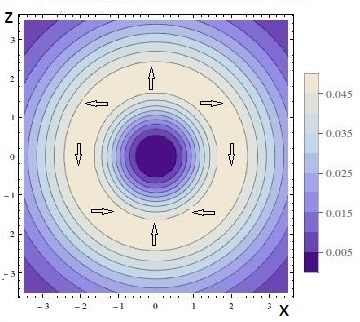}
    \end{picture}
  \caption{
(a) -- Spatial  distribution of electron probability    described by the DE with  the DI for the  states $n=2,\, j=1/2,\,m=\pm1/2,\, \sigma =-$   in the plane $(x,z)$. (b) -- A pattern of two 3D distributions of  probabilities equal to the value 0.02, in the Cartesian coordinate system. (c) -- Charge 'isodensity' lines in the plane $(x,z)$; arrows show spin orientation in the area of the highest charge density for the state $m=1/2$. For the state $m=-1/2,\, \sigma =-$ spins have opposite orientation.  
}
  \label{fig:2}
\end{figure}


\begin{figure}[t]
\vspace{35mm}
\begin{picture}(16,8)
  \includegraphics[width=4cm]{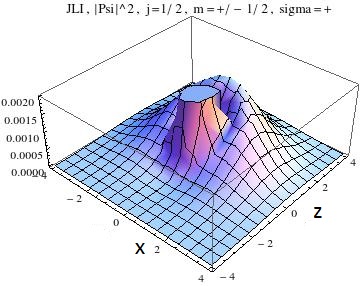}
   \includegraphics[width=3.5cm]{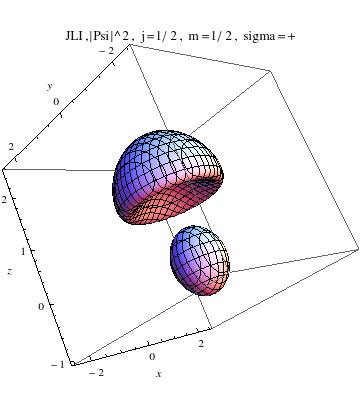}
    \includegraphics[width=4cm]{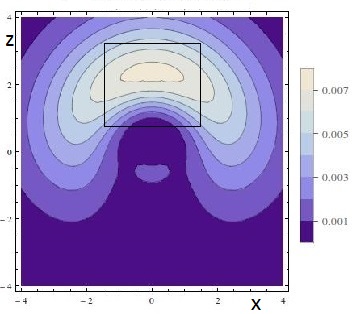}
  \end{picture}
  \caption{
(a) -- Spatial  distribution of electron probability  described by the DE with the JLI for the  states $n=2,\, j=1/2,\,m=1/2,\, \sigma =+$ and  $n=2,\, j=1/2,\,m=-1/2,\, \sigma =-$  in the plane $(z,x)$. (b) -- a pattern of two 3D distributions of  probabilities equal to the value 0.0015, in the Cartesian coordinate system. (c) -- Charge 'isodensity' lines in the plane $(z,x)$. The insert is shown in Fig. 5(a).}
  \label{fig:3}
\end{figure}

\newpage

\begin{figure}[t]
\vspace{40mm}
\begin{picture}(16,8)
  \includegraphics[width=4cm]{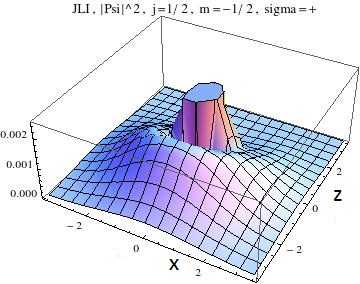}
   \includegraphics[width=4cm]{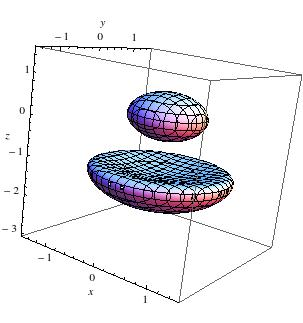}
    \includegraphics[width=4cm]{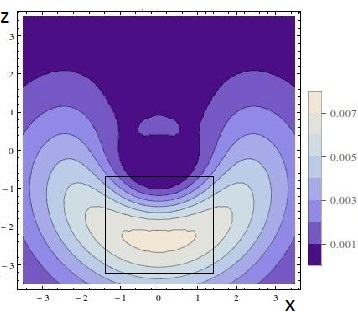}
    \end{picture}
  \caption{
(a) -- Spatial  distribution of electron probability  described by the DE with the JLI for the  states $n=2,\, j=1/2,\,m=-1/2,\, \sigma =+$ and $n=2,\, j=1/2,\,m=1/2,\, \sigma =-$   in the plane $(z,x)$. (b) -- A pattern of two 3D distributions of  probabilities equal to the value 0.0015, in the Cartesian coordinate system. (c) -- Charge 'isodensity' lines in the plane $(z,x)$. The insert showing spin orientation, is represented in Fig. 5(b).}
    \label{fig:4}
\end{figure}


\begin{figure}[t]
\vspace{35mm}
\hspace{25mm}
\begin{picture}(16,8)
  \includegraphics[width=5cm]{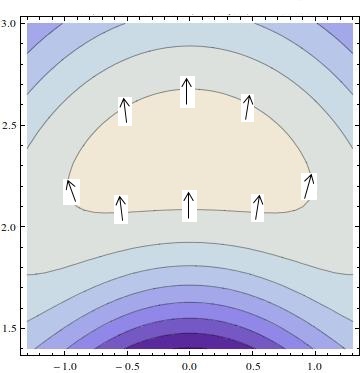}
   \includegraphics[width=5cm]{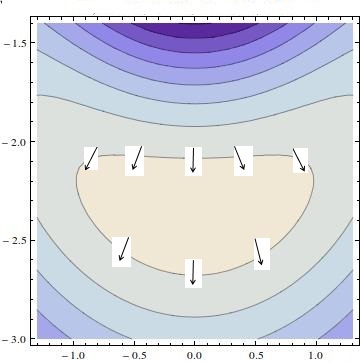}
     \end{picture}
  \caption{ 
 (a) -- The insert of Fig.3 (c), showing spin orientations in the area of the highest electron probability  for the state $n=2,\, j=1/2,\,m=1/2,\, \sigma =+$ for JLI. (b) -- The insert of Fig.4 (c), showing spin orientations in the area of the highest electron probability  for the state  $n=2,\, j=1/2,\,m=-1/2,\, \sigma =+$ for JLI.    
}
  \label{fig:5}
\end{figure}

\begin{figure}[t]
\vspace{4cm}
\begin{picture}(16,8)
  \includegraphics[width=5cm]{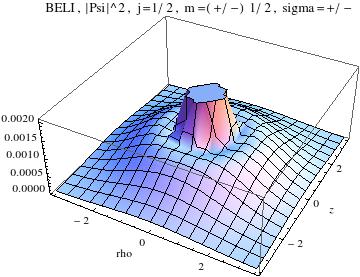}
   \includegraphics[width=4cm]{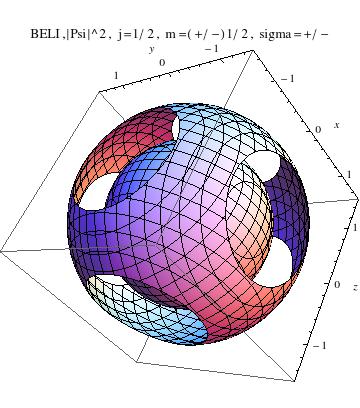}
    \includegraphics[width=5cm]{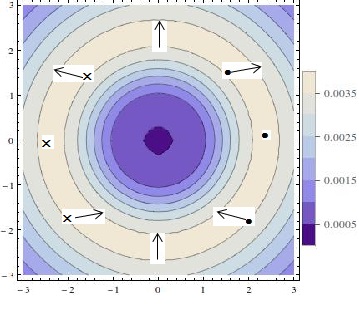}
\end{picture}
 \caption{ (a) -- Spatial  distribution of electron probability  described by the DE with the BELI for the  states $n=2,\, j=1/2,\,m=\pm 1/2,\, \sigma =\pm$   in the plane $(z,\rho)$. (b) -- A pattern of two 3D distributions of  probabilities equal to the value 0.0015, in the Cartesian coordinate system. (c)-- Charge 'isodensity' lines in the plane $(z,x)$; arrows show spin orientation in the area of the highest charge density in the state $m= 1/2,\, \sigma = +$.  Remarkably,  with angle $\varphi $ changing, spins acquire projection out of $(z,x)$ plane  in the up or down direction, indicated by symbols ${ \bullet}$ and $\mathbf{\times}$, respectively. }  
  \label{fig:6}
\end{figure}

\vskip5mm 
{\bf Acknowledgement.} 
\textit{ The work was carried out in the framework
of the budget program KPKVK~6541230 and the scientific program
0117U00236 of the Department of Physics and Astronomy of the
National Academy of Sciences of Ukraine.}

\end{document}